\documentclass[12pt,preprint]{aastex}

\newcommand{\kms}{\mbox{\,km\,s$^{-1}$}}
\newcommand{\Kkms}{\mbox{\,K\,km\,s$^{-1}$}}
\newcommand{\Msun}{\,$M_{\odot}$}

\newcommand{\co}{\mbox{\rmfamily $^{12}$CO}}
\newcommand{\tco}{\mbox{\rmfamily $^{13}$CO}}
\newcommand{\ceont}{\mbox{\rmfamily C$^{18}$O}}
\newcommand{\ceo}{\mbox{\rmfamily C$^{18}$O}\,{(1--0)}}
\newcommand{\Spitzer}{{\it Spitzer}}
\newcommand{\vlsr}{V$_{LSR}$}
\newcommand{\sigmav}{$\sigma_{v}$}
\newcommand{\degree}{$^{\circ}$}
\newcommand{\cms}{${\rm cm}^{-2}$}
\newcommand{\cmc}{${\rm cm}^{-3}$}
\newcommand{\hh}{H$_{2}$}

\newcommand{\tastar}{T$_{A}^{*}$}

\newcommand{\av}{A$_{v}$}
\newcommand{\tsys}{T$_{sys}$}
\newcommand{\tk}{T$_{\mathrm {K}}$}
\newcommand{\tex}{T$_{\mathrm {ex}}$}
\newcommand{\ammonia}{NH$_{3}$}
\newcommand{\ammoniaoneone}{NH$_{3}$ (1,1)}
\newcommand{\ammoniaratio}{NH$_{3}$ (1,1) to (2,2)}
\newcommand{\ammoniatwotwo}{NH$_{3}$ (2,2)}
\newcommand{\ccs}{CCS (2$_{1}$--1$_{0}$)}
\newcommand{\hcfiven}{HC$_{5}$N (9,8)}       % 23963.901
\newcommand{\hcfivennt}{HC$_{5}$N}       % 23963.901
\newcommand{\intint}{$\int$\,\tastar\,$\Delta$V}
\newcommand{\iras}{{\it {IRAS}}}

\slugcomment{}
\shorttitle{Pipe cores: an NH$_{3}$, CCS, and HC$_{5}$N survey}
\shortauthors{Rathborne et al.}

\begin{document}

%%%%%%%%%%%%%%%%%%%%%%%%%%%%%%%%%%%%%%%%%%%%%%%%%%%%%%%%%%%%%%%%%%%%%%%%%%%%%%%%%%%%
\title{The nature of the dense core population in the Pipe Nebula: A survey of NH$_{3}$, CCS, and HC$_{5}$N molecular line emission}
%%%%%%%%%%%%%%%%%%%%%%%%%%%%%%%%%%%%%%%%%%%%%%%%%%%%%%%%%%%%%%%%%%%%%%%%%%%%%%%%%%%%
\author{J. M. Rathborne, C. J. Lada, A. A. Muench}
\affil{Harvard-Smithsonian Center for Astrophysics, 60 Garden Street, Cambridge, MA 02138, USA: jrathborne@cfa.harvard.edu, clada@cfa.harvard.edu, gmuench@cfa.harvard.edu}
\and
\author{J. F. Alves}
\affil{Calar Alto Observatory, Centro Astron\'omico Hispano Alem\'an, c/Jes\'us Durb\'an Rem\'on 2-2, 04004, Almeria, Spain:jalves@caha.es}
\and
\author{M. Lombardi}
\affil{European Southern Observatory, Karl-Schwarzschild-Str. 2, 85748 Garching, Germany: mlombard@eso.org}

%%%%%%%%%%%%%%%%%%%%%%%%%%%%%%%%%%%%%%%%%%%%%%%%%%%%%%%%%%%%%%%%%%%%%%%%%%%%%%%%%%%%
\begin{abstract}
Recent extinction studies of the Pipe Nebula (d=130 pc) reveal many cores
spanning a range in mass from 0.2 to 20.4\,\Msun.  These dense cores
were identified via their high extinction and comprise a starless
population in a very early stage of development. Here we present a
survey of \ammoniaoneone, \ammoniatwotwo, \ccs, and \hcfiven\,emission
toward 46 of these cores. An atlas of the 2MASS extinction maps is
also presented. In total, we detect 63\% of the cores in
\ammoniaoneone, 22\% in \ammoniatwotwo, 28\% in CCS, and 9\% in
\hcfivennt\, emission. We find the cores are associated with dense gas
($\sim$ 10$^{4}$\,\cmc) with 9.5 $\le$ \tk\, $\le$ 17 K.  Compared to
\ceont, we find the \ammonia\, linewidths are systematically narrower,
implying that the \ammonia\, is tracing the dense component of the gas
and that these cores are relatively quiescent. We find no correlation
between core linewidth and size. The derived properties of the Pipe
cores are similar to cores within other low-mass star-forming regions:
the only differences are that the Pipe cores have weaker \ammonia\,
emision and most show no current star formation as evidenced by the
lack of embedded infrared sources. Such weak \ammonia\, emission could
arise due to low column densities and abundances or reduced excitation
due to relatively low core volume densities. Either alternative
implies that the cores are relatively young. Thus, the Pipe cores
represent an excellent sample of dense cores in which to study the
initial conditions for star formation and the earliest stages of core
formation and evolution.
\end{abstract}
%%%%%%%%%%%%%%%%%%%%%%%%%%%%%%%%%%%%%%%%%%%%%%%%%%%%%%%%%%%%%%%%%%%%%%%%%%%%%%%%%%%%
%%%%%%%%%%%%%%%%%%%%%%%%%%%%%%%%%%%%%%%%%%%%%%%%%%%%%%%%%%%%%%%%%%%%%%%%%%%%%%%%%%%%

\section{Introduction}

The Pipe Nebula, at a distance of 130 pc, is one of the closest
molecular cloud complexes in the Galaxy. With a mass of
10$^{4}$\,\Msun, it spans $\sim$ 48 degs$^{2}$ on the sky but
contains little evidence for active star formation. In addition to
tracing the larger-scale molecular gas associated with the Pipe
Nebula, a \co, \tco, and \ceo\, (highly undersampled) survey of the
region \citep{Onishi99} detected a small number of dense cores. The
densest of these, associated with the dark cloud B59, appears to be
the only site within the Pipe Nebula where active star formation is
occurring. Indeed, recent \Spitzer\, observations reveal an embedded
cluster of young stellar objects within B59
\citep{Brooke07}.

\begin{figure}
\begin{center}
\includegraphics[width=0.8\textwidth]{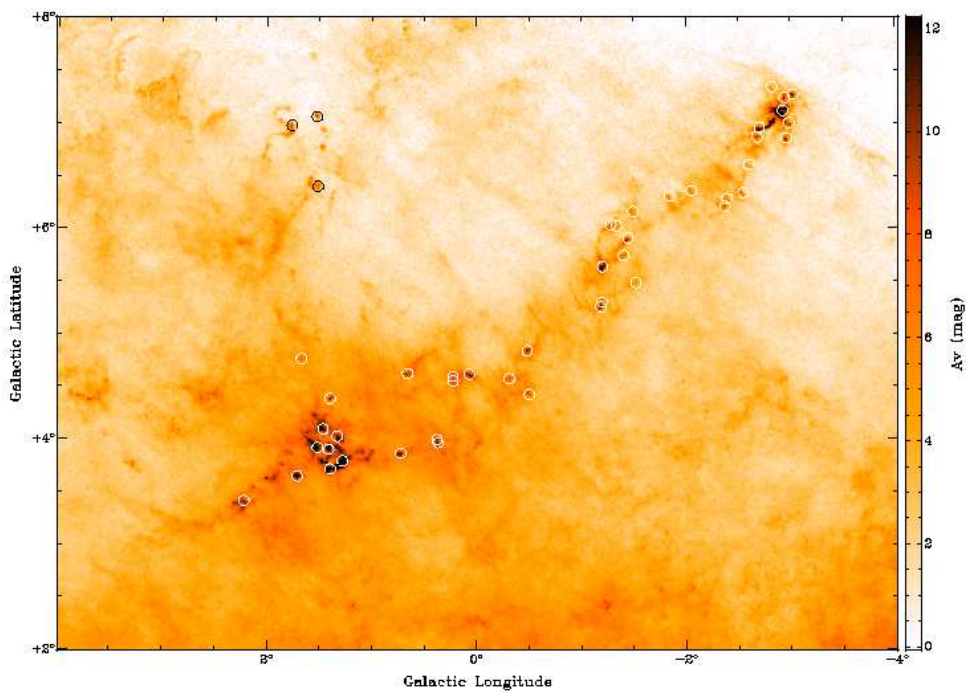}
\caption{\label{pipe}The 2MASS extinction map of the Pipe Nebula \citep{Lombardi06}. This map was 
derived from the JHK photometry of over 4.5 million stars. Many
compact cores are clearly seen in this map in addition to the
larger-scale, more diffuse lower-column density material associated
with the molecular cloud. \cite{Alves07} identified 159 compact cores
within this region.  The 46 cores that comprise our source list are
marked with circles.  These cores were selected to span a range in
peak \av, core mass, and location within the Pipe Nebula. We find that the majority of these
are associated with dense gas.}
\end{center}
\end{figure}

Recent extinction studies of the Pipe Nebula (\citealp{Lombardi06};
Alves, Lombardi, \& Lada 2007) reveal many compact dust cores that
likely represent incipient sites of star formation. These cores were
identified using 2MASS extinction maps derived from the JHK photometry
of over 4.5 million stars (Fig.~\ref{pipe}; \citealp{Lombardi06}). The
individual compact cores were extracted from the larger-scale, more
diffuse lower-column density material associated with the molecular
cloud via a wavelet transform technique (see \citealp{Alves07}). In
total, 159 cores were identified and span a range in peak \av\, and
core mass and are distributed throughout the Pipe Nebula
\citep{Alves07}.  This large sample represents a critical data base
for measuring the initial conditions for star formation and testing
theories of core formation and evolution. Indeed, the dense core mass
function (CMF) generated from these cores reveals a similarity to the
shape of the stellar initial mass function (IMF;
e.g. \citealp{Alves07}) with the profound implication that the stellar
IMF is in fact predetermined by the form of the CMF.  The origin of
the stellar IMF may be, therefore, directly linked to the origin of
dense cores.

While the Pipe Nebula represents an excellent example of a molecular
cloud in a very early stage of star formation, little work has been
done to ascertain the dynamical and evolutionary status of its
cores. Our aim is to study these cores in detail via the combination
of their extinction and molecular line properties; the extinction map
reveals measurements of the core masses and sizes while the
molecular line observations reveal the kinematics and gas temperatures
within the cores.  To this end, we are conducting extensive ground-
and space-based surveys of the cores located within the Pipe Nebula.
The Pipe cores are well suited for studies of core formation and
evolution because they are distributed over a large area, are well
separated and well defined, and represent a complete sample of dense
cores within a single cloud with accurate masses in the mass range of
$\sim$ 0.2 to 20.4\,\Msun.

To begin, we have obtained \ceo\, observations toward a sample of 94
cores distributed throughout the nebula. The details of these
observations and results are presented in \cite{Gus-pipe}. In
brief summary, all cores show strong
\ceont\, emission, have measured linewidths, $\Delta$V, of
$\sim$~0.14 -- 0.61\,\kms, and show no size-linewidth correlation.

Because \ceo\, emission traces all molecular gas along the line of
sight, the measured \ceont\, linewidths will trace both the lower
density intercore material in addition to the dense gas associated
with the core. To isolate the emission from the dense core, we have
also obtained \ammonia\, observations toward a sample of the Pipe
cores and report the results of these observations here. Because
\ammonia\, requires such high densities to be excited ($>$
10$^{4}$\,\cmc), only the densest gas will exhibit significant
\ammonia\, emission. Thus, \ammonia\, observations will reveal the
fraction of cores in the Pipe Nebula that contain dense gas.
Determination of the dense core fraction and how it varies across the
core mass function and with kinematical state will provide crucial
data for understanding the evolutionary status of these objects.

The combination of the extinction and molecular line studies (\ceont\,
and \ammonia), and the implications for core formation and evolution,
is discussed in \cite{Lada-pipe}, the third paper in this series on
the Pipe Nebula.  By accurately measuring the linewidths from the
\ammoniaoneone\, emission in each core, and by combining this with
measurements of the core masses, one can determine the fraction
of cores that are unbound and the core stabilty, which can provide
important clues to their formation and evolution. Indeed,
\cite{Lada-pipe} find that the majority of Pipe cores are pressure confined, 
gravitationally unbound objects whose properties are dominated by
thermal processes in a pressurized medium.

Ammonia has been used extensively to study both high- and low-mass
star-forming regions
(e.g.\,\citealp{Myers83,Benson89,Jijina99}). Because it is the least
depleted species in, and, thus, the most reliable tracer of, cold
dense gas within starless cores, we use \ammonia\, to trace the
densest gas in the Pipe Nebula.  The 18 individual hyperfine
components of the \ammoniaoneone\, transitions can be used to measure
the intrinsic one dimensional velocity dispersion (\sigmav), optical
depth ($\tau$), and excitation temperature (\tex).  Observations of
the \ammoniatwotwo\, transition were also obtained in order to
directly measure the gas temperatures of the brightest cores in the
sample.  Ratios of the \ammonia\, (1,1) and (2,2) lines reveal the
kinetic temperature (\tk) and, therefore, density of the cores.

To probe the very densest gas and to determine the chemical evolution
of the cores, we also obtained \ccs\, and \hcfiven\,
observations. These molecules are found in very dense starless cores
(e.g. \citealp{Benson83,Suzuki92}).  The abundance of carbon-chain
molecules (e.g. CCS and \hcfivennt) compared to \ammonia\, is thought
to trace the chemical evolution of cores; carbon-chain molecules being
more abundant in the very earliest stages in core evolution, compared
to \ammonia\, which is more abundant in the later stages
(e.g. \citealp{Suzuki92}). Thus, the combination of these data can
reveal the youth and star-forming potential of the cores.

In this paper, we present the results of a survey for \ammoniaoneone,
\ammoniatwotwo, \ccs, and \hcfiven\, emission toward a sample of 46 
of the extinction cores identified in the Pipe Nebula. All but one of
these cores has been observed as part of the \ceont\, survey
\citep{Gus-pipe}. We find that most cores are indeed associated with dense
gas. Their measured linewidths are narrow, with several cores having
linewidths comparable to their thermal linewidths. The derived
properties (\tk, intrinsic $\Delta$V, non-thermal $\Delta$V, radius)
of the cores within the Pipe Nebula are similar to cores within other
low-mass star-forming regions: the only difference is that the Pipe
cores typically have weaker \ammonia\, emission and most show no
current star formation.

%%%%%%%%%%%%%%%%%%%%%%%%%%%%%%%%%%%%%%%%%%%%%%%%%%%%%%%%%%%%%%%%%%%%%%%%%%%%%%%%%%%%
\section{Observations and Data Reduction}

The observations were obtained using the 100-m Robert C. Byrd Green
Bank Telescope (GBT) from 2006 October to 2007 February. The
\ammoniaoneone\, \ammoniatwotwo\, \ccs, and \hcfiven\, transitions
were observed simultaneously using the K-band (upper)
receiver. Table~\ref{line-info} lists the frequencies for each of
these transitions.

All spectra were obtained in the frequency switched mode ($\Delta \nu$
= 4 MHz).  The spectrometer was set up with a bandwidth of 12.5~MHz,
which produced a frequency resolution of 1.846~kHz (velocity
resolution of 0.023~\kms).  The GBT beam width at these frequencies is
$\sim$ 30\arcsec.

\begin{figure}
\begin{center}
\includegraphics[width=0.45\textwidth]{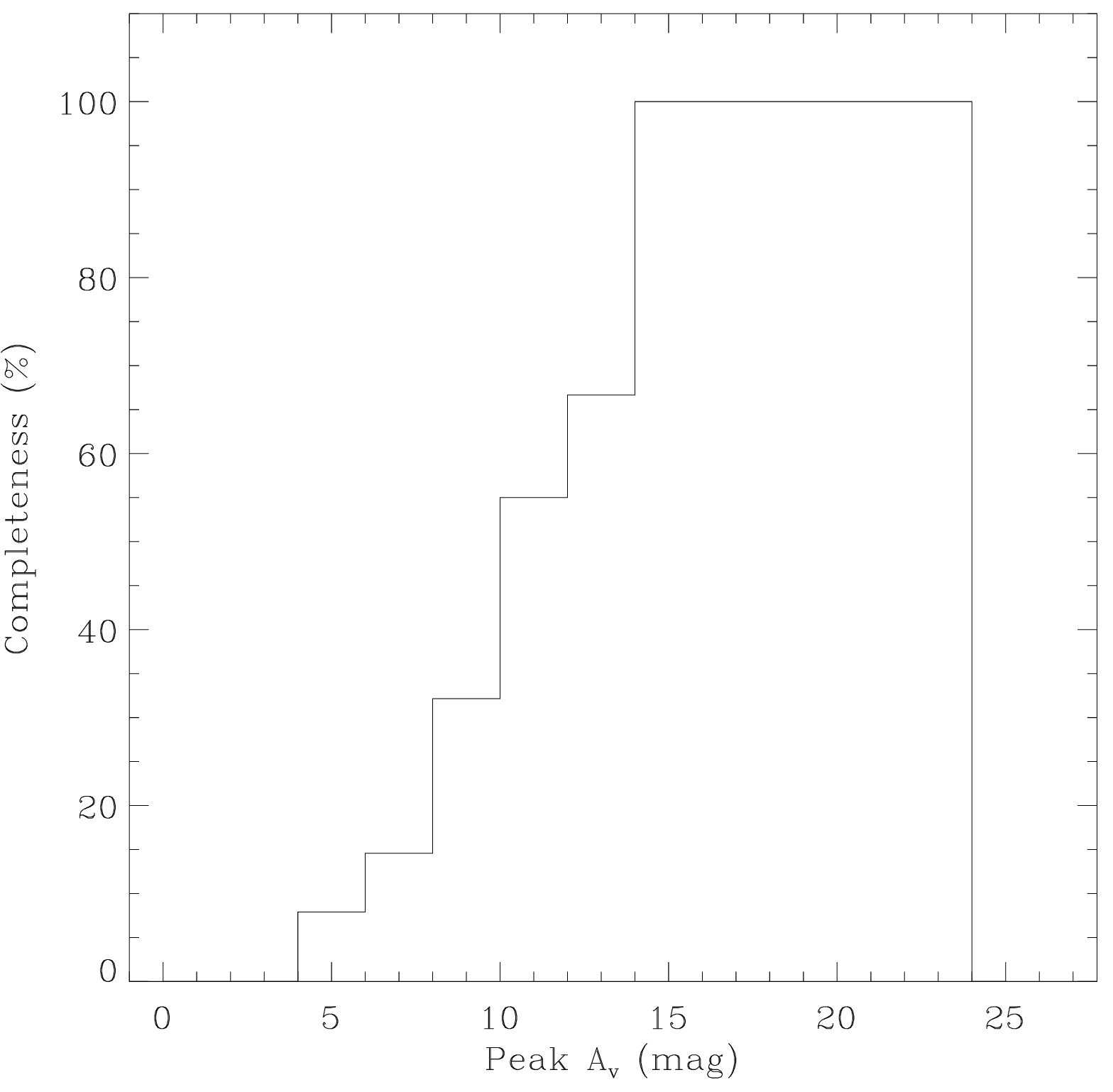}
\includegraphics[width=0.45\textwidth]{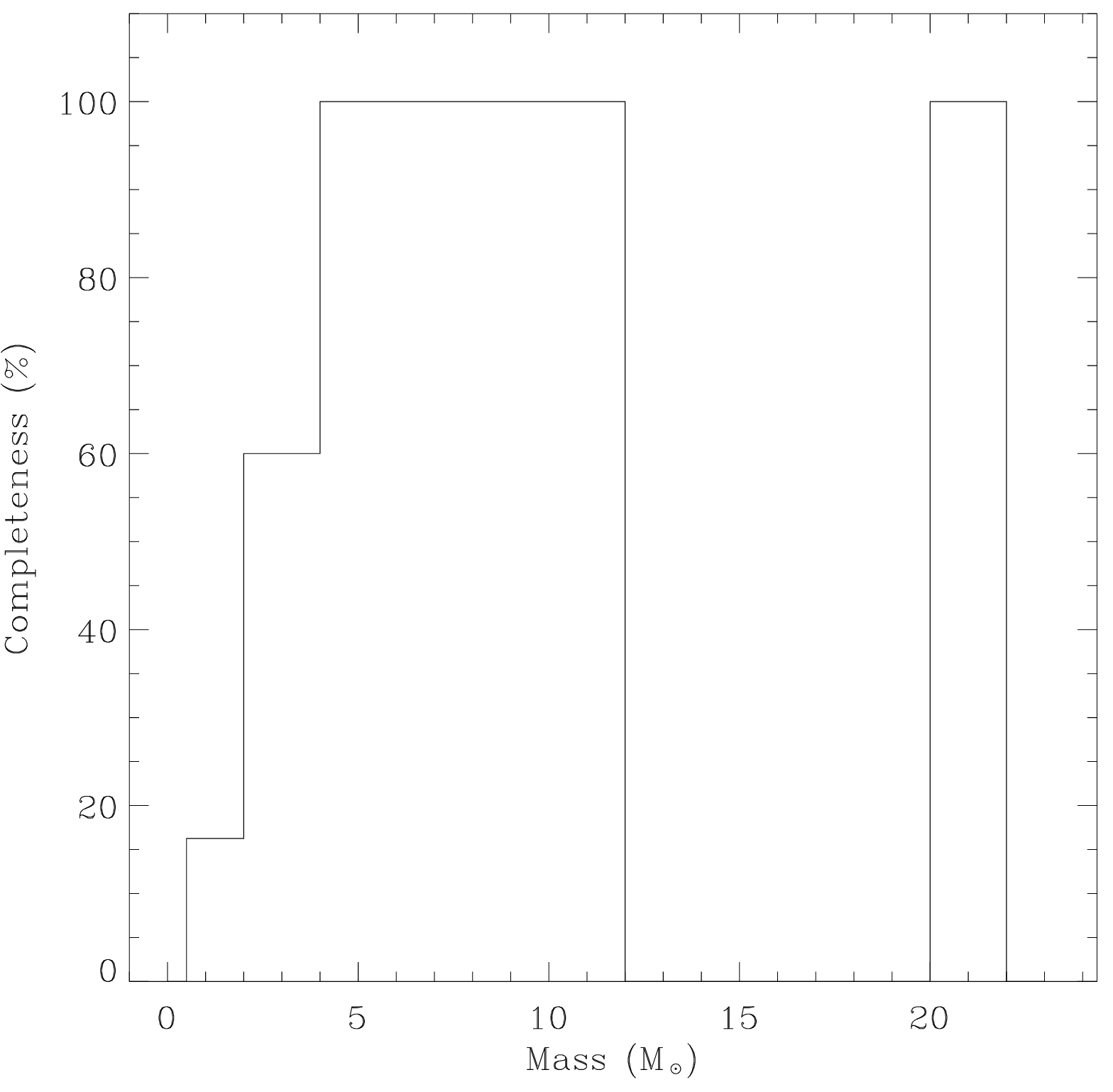}
\caption{\label{completeness}The completeness of the observed sample of cores
given the original catalog of \cite{Alves07} as a function of peak \av\, (left) and core mass (right). All cores 
from the original catalog of \cite{Alves07} with peak \av $>$ 14 magnitudes
and with masses $>$ 4\,\Msun\, were observed.}
\end{center}
\end{figure}

Due to time constraints, a sub-sample of the complete list of cores
identified by \cite{Alves07} were observed. In total, 46 of the 159
cores were observed and were selected to span a range in peak \av,
core mass, and location within the Pipe Nebula. The completeness of
our sample based on the original catalog of \cite{Alves07} is shown in
Figure~\ref{completeness} as a function of both peak \av\, and core
mass. The peak \av\, for the selected cores ranged from 5.3 to 23.7
magnitudes, while their masses range from 0.4 to 20.4\,\Msun; all
cores in the original sample with peak \av\, $>$ 14 magnitudes and
with masses $>$ 4\,\Msun\, were observed.

Spectra were obtained toward the position of peak extinction for each
core (these positions are listed in Table~\ref{table1}). The cores
were originally identified from 2MASS extinction maps ($\Theta_{HPBW}$
$\sim$ 1\arcmin\,; \citealp{Alves07}). Because the coordinates listed
in the catalog of \cite{Alves07} refer to the centroid and not
necessarily the position of peak extinction, the pointing centers for
each core where selected by eye to correspond to the true extinction
peak.  Thus, the coordinates listed here may differ slightly from the
positions reported in the catalog of
\cite{Alves07}.

Toward some of these cores we have also obtained higher-angular
resolution extinction maps using the 3.5 m ESO New Technology
Telescope (NTT; $\Theta_{HPBW}$ $\sim$~15\arcsec;
\citealp{Carlos07}). In many cases, we found sub-structure at
higher-angular resolution. Where possible, therefore, we used the
higher-angular resolution extinction maps to pinpoint the position of
peak extinction. We found that the extinction peaks could differ by as
much as $\sim$ 1\arcmin\, from the quoted positions in the
\cite{Alves07} catalog.  Thus, for the cores in which we have
obtained these higher-angular resolution extinction maps, we use the
coordinates of the peak extinction derived from them.  These cores are
marked with an asterisk in Table~\ref{table1}.

For the majority of cores, four $\sim$ 5 minute integrations were
combined for a total of 18 minutes on source.  Toward the brightest 4
cores only a single 5 minute spectrum was obtained. For three cores
showing very faint \ammoniaoneone\, emission, we combined additional
data for an on source integration time of over 70 minutes.  The
integration times, system temperatures (\tsys), source elevation (El),
and 1 $\sigma$ rms for the spectra are listed in Table~\ref{table1}.
Because the system temperatures for the individual frequencies varied
by $<$ 5\%, we quote a single system temperature and, hence, 1
$\sigma$ rms for each core.

The pointing and focus were checked at the beginning of each observing
session using a nearby, bright pointing source. The pointing
corrections were typically $\sim$8\arcsec\, in Azimuth and
$\sim$1\arcsec\, in Elevation. The flux density scale was determined
from observations of 3C286 (known flux density of 2.38~Jy was taken
from \citealp{Ott94}).

The spectra were converted from the raw data format into fits files
using the SDFITS package. All subsequent data reduction and analysis
was performed using the GBTIDL package. The individual polarizations
for each spectrum were combined and averaged before opacity and
elevation corrections were applied. A first order baseline was removed
from all spectra using user-defined baseline windows.  All quoted
temperatures are \tastar\, (K).

The observations were taken at low elevations ($\sim$~19\degree) where
the pointing model of the GBT is less reliable. As a result, the
temperatures of the observations may be in error, either due to
unintended pointing offsets or pointing drifts during the
integrations~\footnote{The pointing rms during one of the six hour
shifts was 12\arcsec\, in azimuth and 2\arcsec\, in elevation.}.  To
quantify the repeatability of the observations, we obtained a spectrum
toward the core FeSt 1-457 (our core 109) during each observing
session. These data show a variation in peak temperature (\tastar) of
$\sim$ 10\%, while the linewidths show a variation of $\sim$ 3 \%
(these values were estimated by fitting two Gaussians to the blended
central hyperfine components). Thus, while the temperatures listed may
be uncertain, the linewidths and derived parameters (e.g. $\tau$, \tk,
\tex), which are derived from the line ratios and, hence, independent 
of the absolute scale, will be reliable.

%%%%%%%%%%%%%%%%%%%%%%%%%%%%%%%%%%%%%%%%%%%%%%%%%%%%%%%%%%%%%%%%%%%%%%%%%%%%%%%%%%%%
\section{Results}

The \ammoniaoneone, \ammonia\, (2,2), \ccs, and \hcfiven\, spectra for
each core observed in the Pipe Nebula are shown in the Appendix,
together with its 2MASS extinction map. We find many cores show little
or no detectable \ammonia, CCS, nor \hcfivennt\, emission (e.g.
Fig.~36). Many of the remaining cores show only
weak \ammonia\, emission from the central blended hyperfine components
(e.g.\, Fig.~41). A number of cores show weak
\ammonia\, emission, but strong CCS emission
(e.g. Fig.~51).  Only a small fraction of the cores
show strong \ammonia\, emission. Toward these cores, we clearly see
the satellite components in addition to the central blended hyperfine
components (e.g.\,~Figs.~30 and 53).

\subsection{Determining line parameters: \tastar, \sigmav, \vlsr.}

\subsubsection{\ammoniaoneone}

We find 12 cores that show bright \ammoniaoneone\, emission from both
the central and satellite hyperfine components. To determine their
line parameters, we used a forward-fitting routine (for a full
description see \citealp{Rosolowsky07}). This forward-fitting routine
models the spectrum given input physical properties and then optimizes
these values via non-linear least squares minimization. This method is
in contrast to the usual way of determinining physical properties
directly from the \ammoniaratio\, line ratios. The advantage of this
method is that the uncertainties for the physical parameters can be
determined in addition to the covariance among the parameters.

In this model, the gas is assumed to have a slab geometry with uniform
properties and that the velocity dispersions and excitation
temperatures of the transitions are the same, i.e. the gas is in Local
Thermodynamic Equilibrium (LTE). Assuming the beam filling factor is
unity, the spectrum is entirely determined by five parameteres:
\tk, \tex, $\tau$, \sigmav, and \vlsr. Because the exinction cores
are always several times larger than the GBT beam, the assumption
that the beam filling factor is unity is valid. 

The spectra are easily modelled using this approach when the opacity
is high. In the case of low optical depth, however, the parameters
\tex\, and $\tau$ become degenerate and it is not possible to solve
for them independently. Instead, the fit is optimized for the free
parameter {\mbox{$\gamma\equiv$(\tex\,$-$T$_{bg}$)$\tau$}} and, thus,
in this case, the spectrum is determined by four free parameters: \tk,
$\gamma$,
\sigmav, and \vlsr.

Table~\ref{line-parameters-1} lists the measured \tastar, \vlsr,
\sigmav, and integrated intensity for all cores showing \ammoniaoneone\,
emission with detections $>$ 3\,$\sigma$. For the 12 bright cores, the
\vlsr\, and \sigmav\, were determined using the forward-fitting routine discussed
above (these cores are marked with an asterisk in
Table~\ref{line-parameters-1}). The quoted \tastar\, in these cases
were measured from a single Gaussian fit to the central blended component. The
integrated intensity was calculated by summing the emission across
the spectrum (from V=$-$23 to 30\,\kms).

Toward the remaining 34 cores observed, we find 17 cores with weak
\ammoniaoneone\, detections.  Because the emission is considerably
weaker in these cases, only the central blended components were
detected ($>$ 3$\sigma$).  Thus, to characterize this emission, we fit
a single Gaussian profile to the observed emission. The \tastar,
\vlsr, and integrated intensity {\mbox {(I = \intint)}} measured from
these fits are included within Table~\ref{line-parameters-1}.

Because the central component is a blend of 8 individual hyperfine
components, the velocity dispersion determined from fitting a single
Gaussian profile to the emission will be overestimated. To constrain
theories of core formation and evolution, we require accurate
estimates of the core velocity dispersions. Thus, it was necessary to
determine the amount by which this measurement of the velocity
dispersion was overestimated.

To derive the intrinsic velocity dispersion in the cases where we
could not fit the spectra with all 18 hyperfine components, we used a
conversion factor calculated from synthetic spectra. The synthetic
spectra were generated to cover a range in \sigmav, \tex, and
$\tau$. To each of these spectra, we fit a single Gaussian profile to
the central blended components and then compared the measured velocity
dispersion to that which was used to generate the
spectrum. 

To determine the correction factor for all measured velocity
dispersions we interpolated the relation to produce a curve converting
the measured velocity dispersion to that of the intrinsic velocity
dispersion (our derived relation is similar to that modelled by
\citealp{Barranco98}). Thus, the correction factor was determined
using this method and then applied to each velocity dispersion
measured from a single Gaussian fit to the central blended hyperfine
components (17 cores). These corrected velocity dispersions are quoted in
Table~\ref{line-parameters-1}.

In total, toward the 46 cores observed, \ammoniaoneone\, emission was
detected ($>$~3$\sigma$) in 29 cases ($\sim$~63\%). \ammoniaoneone\,
emission was detected toward cores spanning the full range of peak
\av\, and core mass within our sample. We find no clear correlation between
the measured line parameters and either peak \av\, or core mass. Most
($\sim$ 86 \%) of the cores have \tastar~$<$~1~K. The \vlsr\, measured
for the cores range from 3 to 6\,\kms\, consistent with the measured
\vlsr\, of the molecular gas associated with Pipe Nebula (as traced by
\ceont; \citealp{Onishi99,Gus-pipe}). This confirms that these cores are indeed
located within the large-scale molecular cloud associated with the
Pipe Nebula and not either background or foreground objects.  The
measured velocity dispersions range from $\sim$ 0.05 to 0.3\,\kms.

\subsubsection{\ammoniatwotwo, CCS, and \hcfivennt}
 
For the \ammoniatwotwo, CCS, and \hcfivennt\, emission, we fit a
single Gaussian profile to all spectra with a $>$ 3 $\sigma$
detection.  Although the \ammoniatwotwo\, emission comprise 21
individual hyperfine components, in all cases only the central
12 blended components were detected. Tables~\ref{line-parameters-1} and
\ref{line-parameters-2} list the peak \tastar, \vlsr, \sigmav, 
and integrated intensity obtained from these fits.

Toward the 46 cores observed, \ammoniatwotwo\, emission was detected
in 10 cores ($\sim$~22\%), CCS emission in 13 cores ($\sim$~28\%), and
\hcfivennt\, emission in 4 cores ($\sim$~9\%).  

We find \ammoniatwotwo\, detections toward cores spanning the range of
peak \av\, and core mass, with the most massive core showing the
brightest \ammoniatwotwo\, emission. In contrast, we only detect CCS
emission for cores with peak \av\, $>$ 8 magnitudes and \hcfivennt\,
emission toward cores with peak \av\, $>$ 12 magnitudes. This is
expected given that the critical densities of the CCS and \hcfivennt\,
transitions ($\sim$ 10$^{5}$\,\cmc) are higher than for \ammonia\,
($\sim$ 10$^{4}$\,\cmc) and that there may be differences in their
relative abundances due to evolutionary effects.

\subsection{Physical properties: \tk, $\tau$, \tex, n(\hh), column densities}
\label{ammonia-nhh}

\begin{figure}
\begin{center}
\includegraphics[angle=90,width=0.45\textwidth]{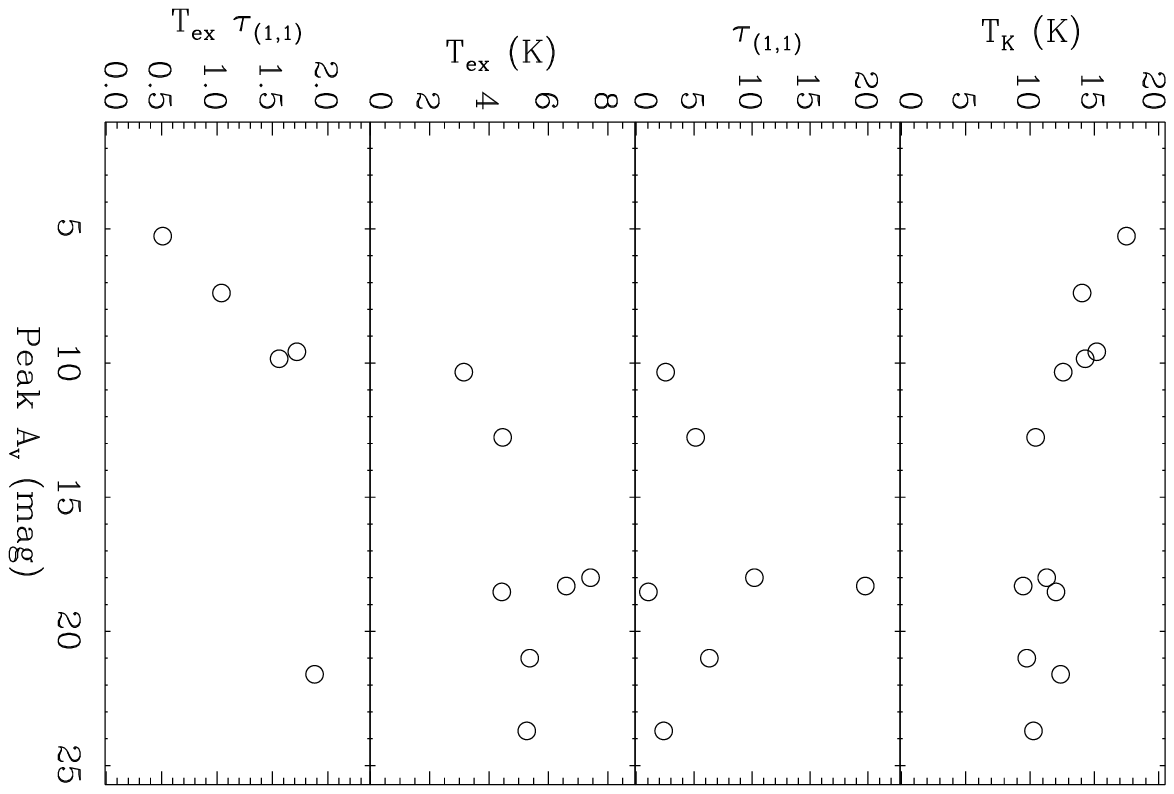}
\includegraphics[angle=90,width=0.45\textwidth]{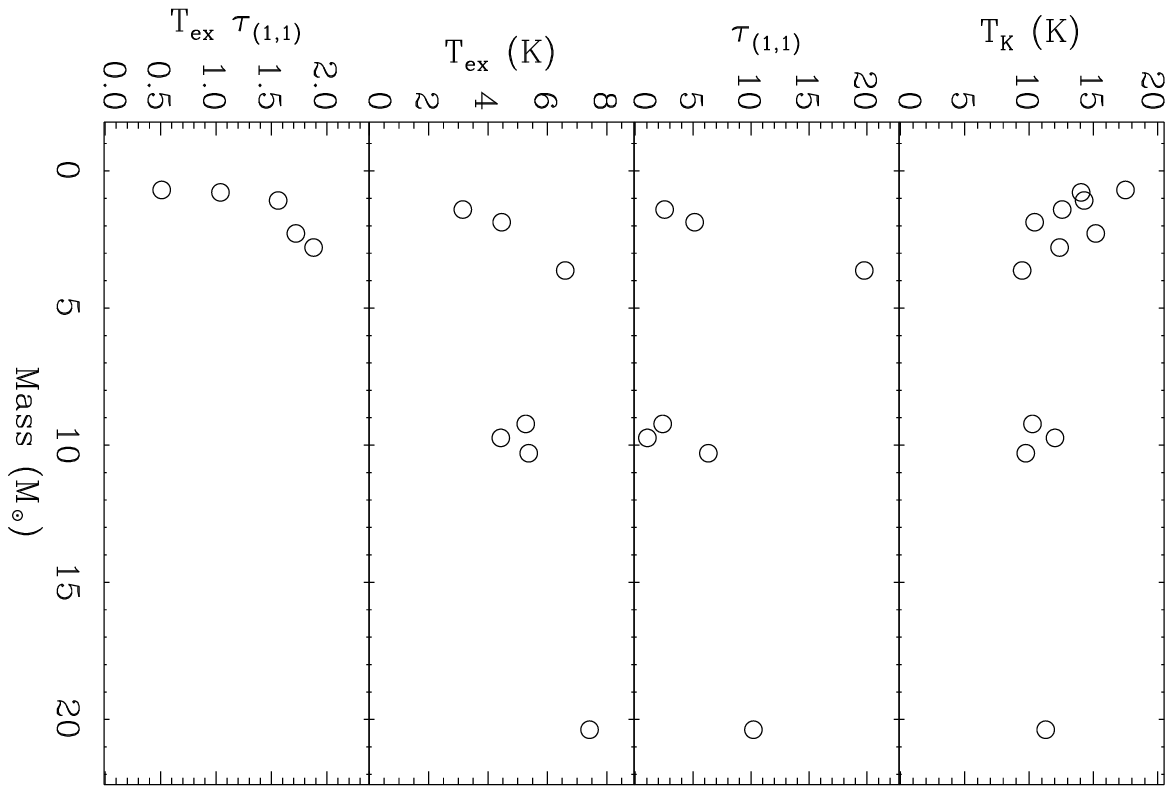}
\caption{\label{phys-av-mass}The derived parameters, \tk, $\tau$, and \tex, output from 
the forward-fitting modelling of the cores showing strong \ammonia\,
emission as a function of peak \av\, (left) and core mass (right). For
cores with low opacities, values of $\tau$ and \tex\, can not be
determined independently. The lower panels shows the values of the
parameter $\tau$\tex\, which is used to characterize the emission in
the case of low optical depths.}
\end{center}
\end{figure}

The forward-fitting model that was used to derive the \sigmav\,  and
\vlsr\, for the cores showing bright \ammonia\, emission, also
measures the physical properties of the gas. Figure~\ref{phys-av-mass}
shows the measured \tk, $\tau$, and \tex\, for the brightest cores as
a function of the peak \av\, and core mass.

For the 12 cores showing bright \ammonia\, emission we find that the
mean \tk\, is $\sim$~12~$\pm$~2~K. However, for the cores with
masses $>$~5\,\Msun\, we find \tk\, of $\sim$~10 $\pm$ 1 K, while
cores with masses $<$~5\,\Msun\, have \tk\, of $\sim$~13 $\pm$ 3 K
(the errors quoted here correspond to the standard deviation). 
Because \tk\, is measured from the ratios of the \ammonia\, (1,1) to
(2,2) lines, this calculation is independent of assumptions about the beam filling factor.
The measured opacities for all the cores range from $\sim$~2 to 10,
with a single core having a $\tau$ of $\sim$~20 (this is FeSt 1-457,
our core 109). The typical \tex\, was $\sim$~5~$\pm$~1~K.

In the low optical depth regime, the values of $\tau$ and \tex\,
become degenerate, so the fitting routine optimizes the value of the
parameter $\gamma$ ($\propto \tau$\tex).  For the 5 cores that were
found to lie in the optically thin regime, the values of $\tau$\tex\,
are plotted in the lower panel of Figure~\ref{phys-av-mass}.  These
cores all have masses $<$ 5\,\Msun\, and calculated values of
$\tau$\tex\, ranging from 0.5 to 2.

To verify the values of \tk\, output from the forward-fitting model,
we have also calculated the rotation temperature, T$_{R}$, which is
related to \tk\, via \citep{Swift05}

\[ T_{R} = T_{K} \{ 1 + \frac{T_{k}}{T_{0}} ln \left[ 1 + 0.6 e^{-15.7/T_{k}} \right] \}^{-1} \]

\noindent where T$_{0}$ is the energy gap between the two levels (41.5 K). 
Using this relation we calculate T$_{R}$ for the 7 cores with measured
$\tau$ values to be $\sim$ 10.2 $\pm$ 0.9 K. The rotation temperature
can also be determined directly from the ratios of the peak
\tastar\, for the \ammoniaoneone\, and (2,2) emission via the expression \citep{Ho83}

\[ T_{R} = \frac{-41.5}{ln \left[ \frac{-0.282}{\tau} ln \{ 1 - \frac{T_{A}^{*}(2,2)}{T_{A}^{*}(1,1)} (1 - e^{-\tau}) \} \right] }\]

\noindent where $\tau$ is the opacity listed in Table~\ref{fitspec} and 
\tastar (1,1) and \tastar (2,2) are the measured peak
temperatures listed in Table~\ref{line-parameters-1}. Using this
expression, we derive T$_{R}$ of $\sim$ 9.5 $\pm$ 1.8 K. Thus, while
we find that the mean T$_{R}$ measured directly from the line ratios
is slightly lower than the T$_{R}$ calculated from the forward-fitting
routine, they are consistent within the errors especially considering
the peak \tastar\, are measured from a single Gaussian fit to the
central blended components. 

Moreover, we find that the measured T$_{R}$ is comparable to \tk\,
within the errors. \cite{Martin82} show that while T$_{R}$ is, in
general, lower than \tk, T$_{R}$ is a good indicator of the \tk\, for
kinetic temperatures $<$ 20~K. However, T$_{R}$ significantly
underestimates \tk\, for higher temperatures. Because our measured
\tk\, values are always lower than 20~K, the similarity of the derived
T$_{R}$ and \tk\, values is expected.
												
With the determination of \tex\, and \tk, the \hh\, volume density,
n(\hh), can be calculated via the expression \citep{Ho83}

\[ n(H_{2}) = \frac{A}{C} \left[ \frac{J_{\nu}(T_{ex}) - J_{\nu}(T_{bg})}{J_{\nu}(T_{k}) - J_{\nu}(T_{ex})} \right] \left[ 1 + \frac{J_{\nu}(T_{k})}{h\nu/k} \right]\]% \hspace{0.8cm} ({\rm cm^{-3}})\]

\noindent where A is the Einstein A coefficient (1.67$\times$10$^{-7}$
s$^{-1}$; \citealp{Ho83}), C is the rate coefficient for collision de-excitation 
(8.5$\times$10$^{-11}$~cm$^{3}$\,s$^{-1}$;
\citealp{Danby88}) and $J_{\nu}(T)$ is the Planck function at the given
temperature, where, $J_{\nu}(T) = h\nu/k\,({\rm
{exp}}(h\nu/kT)-1)^{-1}$. Table~\ref{column-densities} lists the
n(\hh) values for the cores with \ammonia\, emission sufficiently
bright enough to reliably calculate both \tk\, and \tex. For these 7
cores we find the n(\hh) values ranged from 0.1 to
2.5~$\times~$10$^{4}$~\cmc.  The measured n(\hh) derived from the
extinction maps for these 7 cores range from 0.5 to
1.3~$\times$~10$^{4}$\,\cmc. In detail, we find that the n(\hh) derived
from the \ammonia\, observations reveal densities typically twice those
estimated from the extinction maps.  Because the n(\hh) derived from
the extinction maps represents the mean n(\hh) within the core, we
expect these values to be lower than the n(\hh) calculated from the
\ammonia\, emission.

With the measured opacity and assuming LTE conditions, we calculate
the \ammonia\, column density following the method of \cite{Olano88}
via the expression

\[ N({\rm{NH_{3}}}) = 1.3773 \times 10^{13}\, Z(T_{k})\, \tau\, \Delta V \, T_{ex}\]% \hspace{0.8cm} ({\rm cm^{-2}})\]

\noindent where Z(T$_{k}$) is the partition function (4.417 for \tk~=~10~K;
\citealp{Olano88}). For this calculation we assume \tk = 10 K and use
the measured values of $\tau$, \tex, and $\Delta$V (where $\Delta$V =
2.35\,\sigmav) for each core.  Table~\ref{column-densities} lists the
values obtained for the cores in which we could determine $\tau$ and
\tex\, (or $\tau$\tex).  For these cores, the N(\ammonia) was
typically $\sim$~4~$\times$~10$^{14}$~\cms.

For CCS, \hcfivennt, and \ceont\, we assume the emission is optically
thin and calculate the column density via 

\[ N =\frac{ 3 k^{2}}{8 \pi^{3} h B}\frac{1}{\mu^{2}(J_{1} + 1) \nu}\frac{T_{ex}J_{\nu}(T_{ex})\,{\rm exp}(E_{u}/kT_{ex})}{J_{\nu}(T_{ex}) - J_{v}(T_{bg})}\,\int T_{b} dV\]

\noindent where J$_{1}$ is the rotational quantum number of the lower level,
T$_{bg}$ is the background temperature (2.7 K), E$_{u}$ is the energy
of the upper level of the transition, $\mu$ is the dipole moment, and
B is the rotational constant (Table~\ref{line-info} lists the values
used for these parameters). The quantity $\int$T$_{b}\Delta$V is
simply the integrated intensity listed from
Table~\ref{line-parameters-2} (with the appropriate beam efficiency
conversion from \tastar\, to T$_{b}$, which is 0.58 for the GBT at
these frequencies). For this calculation we assume LTE conditions and
that \tex=10~K.  We find typical N(CCS) of
$\sim$\,5$\times$10$^{12}$~\cms, N(\hcfivennt) of
$\sim$\,3$\times$10$^{12}$~\cms, and N(\ceont) of $\sim$ 1 $\times$
10$^{15}$\,\cms\, (see Table~\ref{column-densities}). The derived
column densities will be under estimated if the excitation temperature
is lower than the assumed value of 10\,K.

For the 17 cores that were detected in \ammoniaoneone\, but
showed no \ammoniatwotwo\, emission, \tk, $\tau$\, or
\tex\, could not be determined using the forward-fitting routine, 
due to its requirement for emission lines in both spectra. In these
cases we also calculate the \ammonia\, column density assuming that
the emission is optically thin using the above expression and that
\tex=5~K. This value of \tex\, was the typical measured \tex\, for the 7 cores for which it
could be determined from the \ammonia\, hyperfine components. For
these 17 cores we find the typical N(\ammonia) was $\sim$ 3 $\times$
10$^{13}$\,\cms.

%%%%%%%%%%%%%%%%%%%%%%%%%%%%%%%%%%%%%%%%%%%%%%%%%%%%%%%%%%%%%%%%%%%%%%%%%%%%%%%%%%%%
\section{Discussion}

\subsection{The dense core detection rate}

\begin{figure}
\begin{center}
\includegraphics[angle=90,width=0.45\textwidth,clip=true]{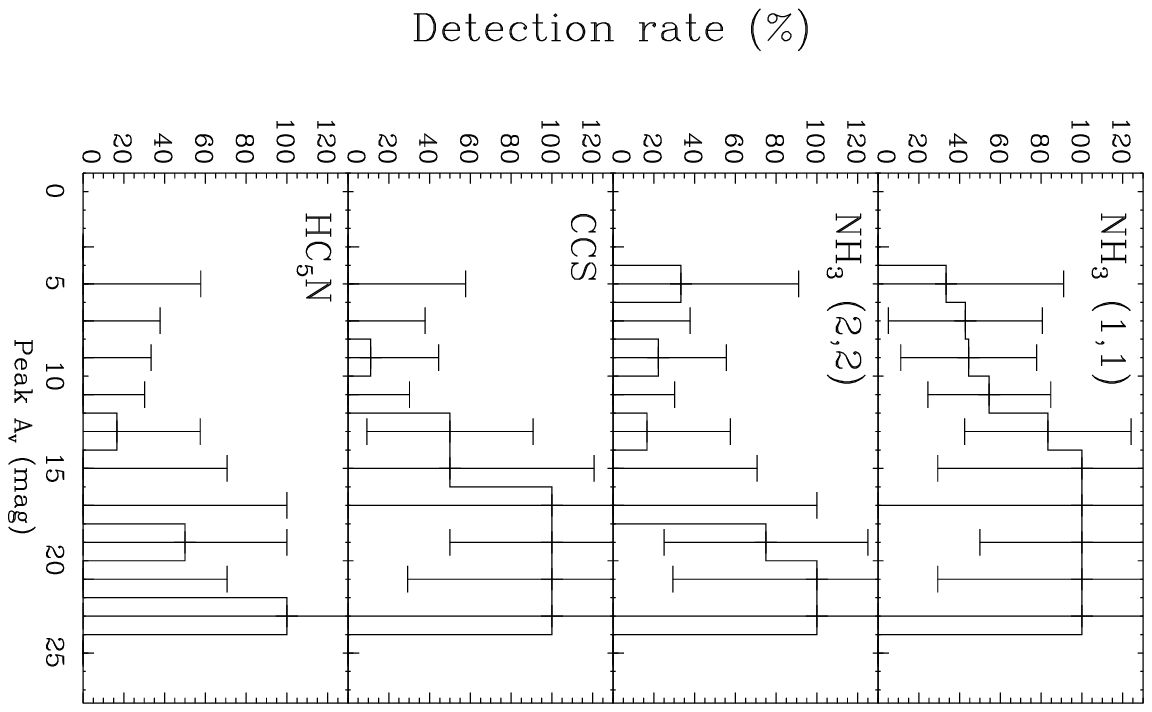}
\includegraphics[angle=90,width=0.45\textwidth,clip=true]{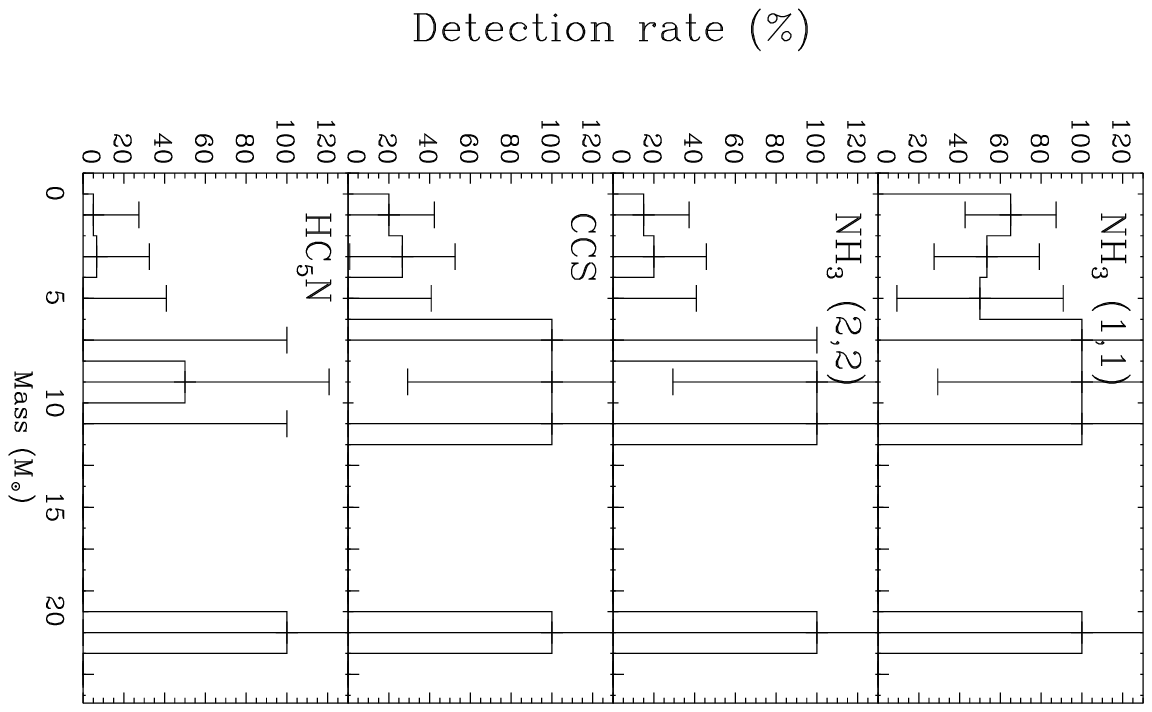}
\caption{\label{detection-rates}The detection rate for each molecular transition as a 
function of peak \av\, (left) and core mass (right).  We detect all
\ammoniaoneone\, emission toward cores with a peak \av $>$ 15
magnitudes and with masses $>$ 11 \,\Msun.}
\end{center}
\end{figure}

Figure~\ref{detection-rates} shows the detection rate for the
\ammoniaoneone, \ammoniatwotwo, CCS, and \hcfivennt\, transitions
as a function of peak \av\, and core mass.  Not surprisingly, the
detection rate is higher toward those cores with higher peak \av\, and
higher mass; in the case of \ammoniaoneone\, we detected emission
toward all cores with a peak \av $>$ 15 magnitudes and with
masses~$>$~11\,\Msun. 

\begin{figure}
\begin{center}
\includegraphics[width=0.6\textwidth,clip=true]{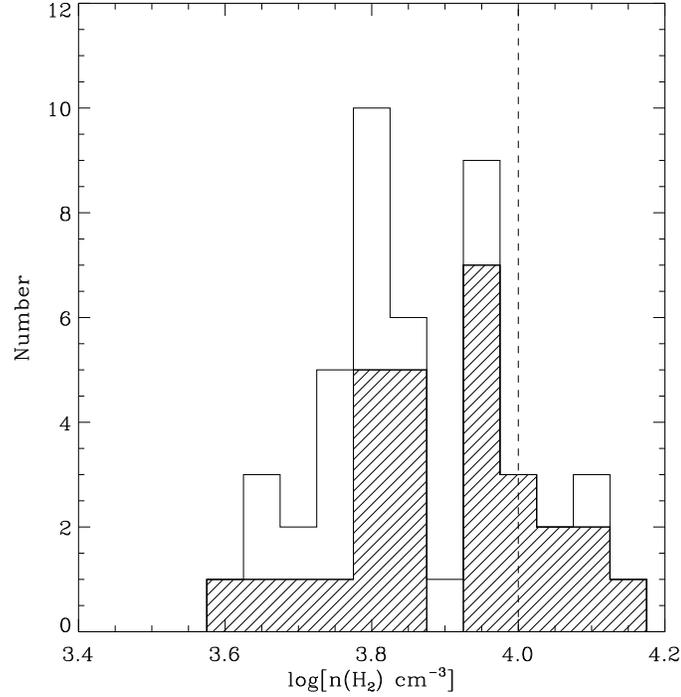}
\caption{\label{density}The mean volume \hh\, density, n(\hh), calculated from the visual 
extinction measurements for the 46 cores within our sample (open
histogram). The diagonal lines mark the n(\hh) measured from the
extinction maps for the cores detected in \ammoniaoneone.  The dotted
line marks the critical density of the
\ammoniaoneone\, transition. We find $\sim$ 17 \% of the cores have
mean volume densities greater than the critical density for
\ammonia. Our higher measured \ammoniaoneone\, detection rate (63\%)
is expected considering that \ammoniaoneone\, traces the densest inner
regions within the cores. This is in contrast to the n(\hh) measured
from the extinction maps which corresponds to the mean volume density
within the larger extinction core.}
\end{center}
\end{figure}

Figure~\ref{density} shows the histogram of the mean \hh\, volume
densities, n(\hh), for the 46 cores surveyed in \ammonia, obtained
from the visual extinction measurements. For this calculation, a
spherical volume with radius equal to half the measured size of the
core was assumed (as listed in \citealp{Alves07}). The masses were
calculated from the background subtracted extinction map (see
\citealp{Lada-pipe} and Table~\ref{table1}).  We find the n(\hh) for the 
complete sample of 46 cores range from 0.4 to 1.3 $\times$
10$^{4}$\,\cmc.  Based on the critical density of \ammonia\, (n$_{c}$
$\sim$ 10$^{4}$\,\cmc), we crudely estimate that the detection rate
should be $\sim$ 17\%.  Because these volume densities were derived
from the extinction maps, they correspond to the mean volume densities
within the cores. The fact that we measure a higher detection rate
from our observations ($\sim$ 63\%) is not surprising given that
\ammonia\, traces the densest inner parts of a compact core and that
the cores most likely have inwardly increasing density gradients.

The detection rate for the \ammoniatwotwo, CCS, and \hcfivennt\,
transitions are much lower compared to \ammoniaoneone\, likely due to
the combination of the fact that higher densities are typically needed
for their excitation and that differences in their chemical
evolutionary stages will effect their relative abundances. In the case
of CCS, its relative distribution to \ammonia\, within the cores may
differ significantly. In many dense, starless cores the distribution
of \ammonia\, and CCS emission is noticably different; the \ammonia\,
emission is typically centrally concentrated, while the CCS emission
arises from a ring-like structure around the \ammonia\, core
(e.g. \citealp{Benson83,Velusamy95,Kuiper96,Hirota02}). Indeed,
morphological differences in the distributions of various depleted
molecules, typically carbon bearing species, is clearly seen toward
the B68 (our core 101) and FeSt 1-457 (our core 109;
\citealp{Bergin02,Aguti07}).

\begin{figure}
\begin{center}
\includegraphics[angle=90,width=0.45\textwidth,clip=true]{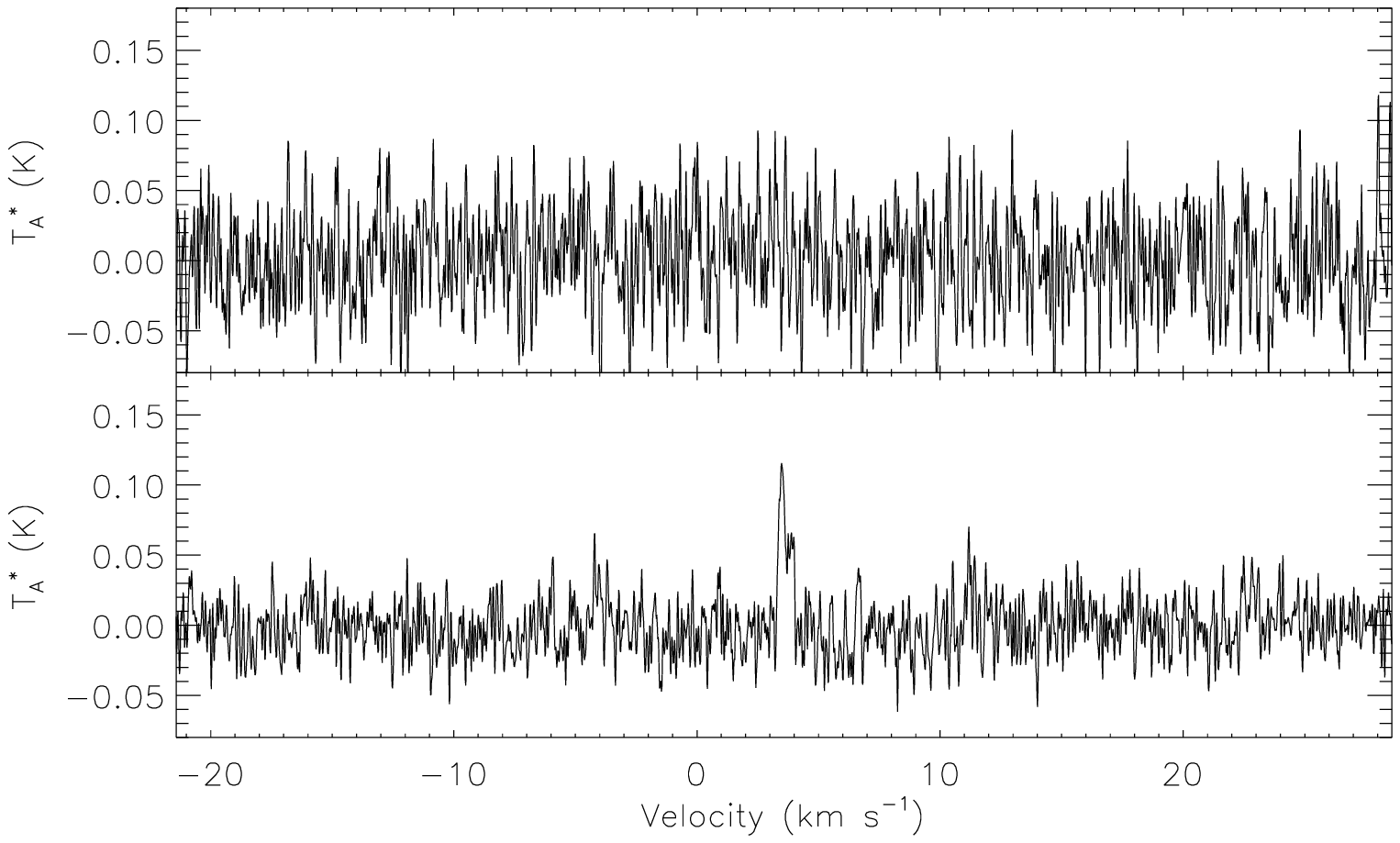}
\includegraphics[angle=90,width=0.45\textwidth,clip=true]{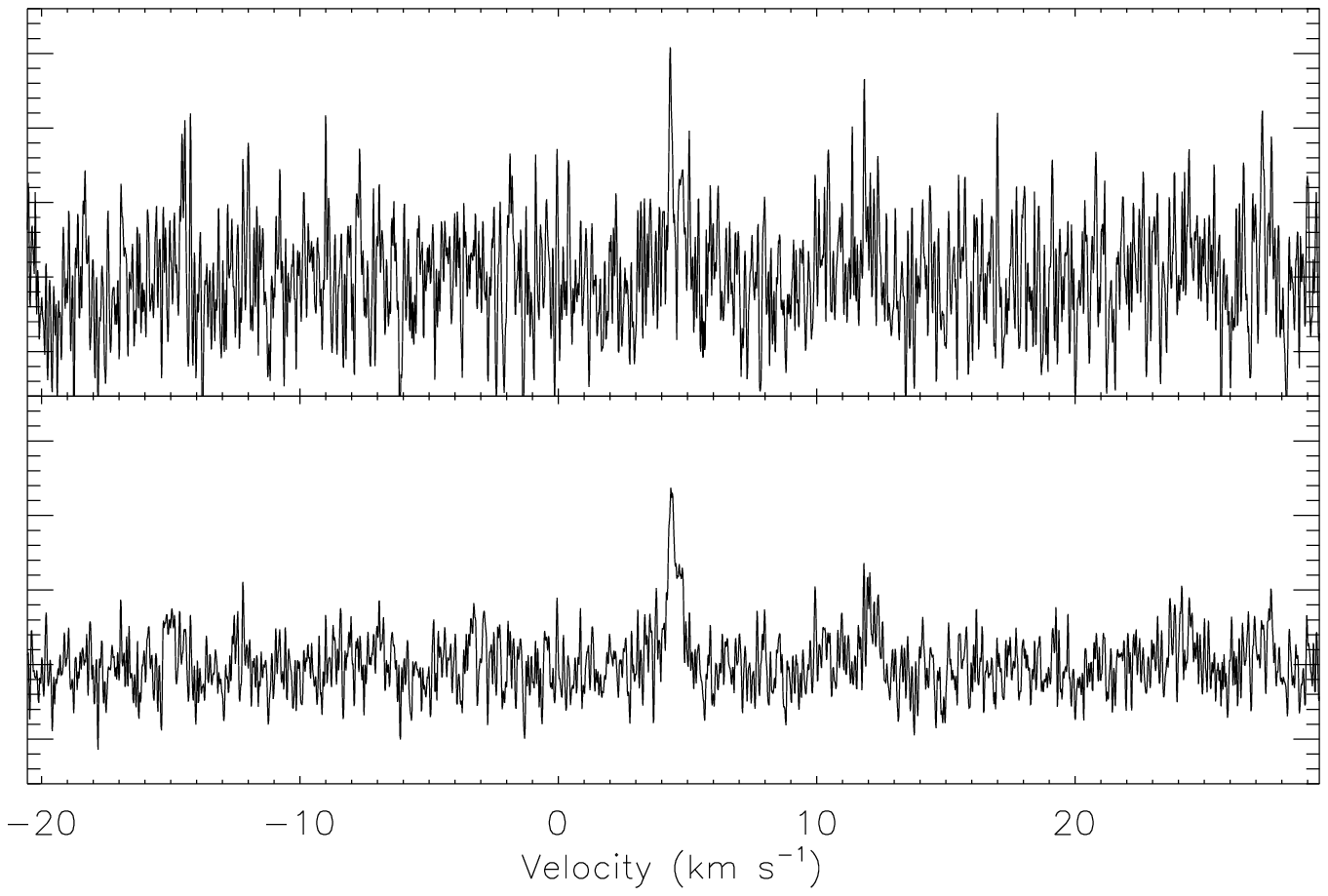}
\caption{\label{comparisons}The comparison between \ammoniaoneone\, spectra obtained with
differing integration times (top: t $\sim$ 18 minutes, bottom: t $\sim$
74 minutes) for two cores (23 and 89, left and right respectively). The spectra toward the
majority of cores were obtained with an integration time of $\sim$ 18 minutes (c.f. top panels).
In these two cases, therefore, we would not have detected \ammoniaoneone\, emission. Thus, 
 the non-detection of
\ammoniaoneone\, emission may simply be due to the short integration
times. With longer integration times per point the detection
rate for these cores would likely increase.}
\end{center}
\end{figure}

The non-detection of \ammonia\, emission toward many of the cores
could arise because of the short integration time per position or
because the spectra were not obtained toward the true density peak of
the core. Because we are primarily interested in the \ammonia\, line
ratios and velocity dispersions, our results will not be significantly
affected by the fact that we may not be pinpointing the peak of the
\ammonia\, emission. This will, however, effect the detection rate as
we may be missing some emission that falls below our sensitivity
limits given the short integration time per point. This is clearly
seen in Figure~\ref{comparisons} which compares an 18 minute
integration to a 74 minute integration toward the same position in two
cores (cores 23 and 89). The spectra obtained with the shorter
integration time are noiser and, thus, the emission in these cases
would not be considered a 3$\sigma$ detection (upper panels). The
longer integrations (lower panels) reveal that indeed \ammoniaoneone\,
emission does arise from within these cores, albiet the emission is
considerably weaker. Given longer integration times per point, the
detection rate for these cores would likely increase.

Because only a single spectrum was obtained toward each core, maps are
necessary to determine if many of the non-detections may arise due to
either inaccurate coordinates from the extinction maps or pointing
errors due to the low observing elevations.  For 36 of the cores,
higher-angular resolution exinction maps, generated from NTT images
(15\arcsec\, angular resolution; \citealp{Carlos07}), were used to
locate the position of peak extinction. For the remaining 13 cores, we
used the lower-angular resolution 2MASS extinction maps (1\arcmin\,
angular resolution; \citealp{Alves07}). When comparing the positions
of peak extinction between the two maps, we found that the position of
peak extinction in the higher-angular resolution maps could differ
from that obtained from the lower-angular resolution maps by $\sim$
1\arcmin\, ($\sim$ 2 GBT beams at these frequencies). In some cases,
the extinction cores identified from the 2MASS maps actually consisted
of several sub-structures in the higher-angular resolution maps. Thus,
the positions obtained for the 13 cores from the 2MASS exinction map
may not reflect the true position of peak extinction, or density, in
the core at the angular scale of the GBT beam. However, because the
cores are resolved in the 2MASS extinction map (typical sizes of
6\arcmin), and with respect to the GBT beam, the positions at which
the spectra were obtained will still lie within each of the
cores. They may not, therefore, necessarily correspond to the true
peak of the extinction, density, nor the peak of the \ammonia\,
emission. This will be most obvious if the molecular line emission
arises from a very compact, dense core.  Thus, the non detection in
these cases may simply arise due to the fact that we were not pointing
at the best position within the core and that the dense cores may be
compact relative to the extinction cores. With only a single spectrum
toward the cores it is difficult to rule out potential pointing
offsets. Obtaining \ammonia\, and CCS maps toward each of the cores
will reveal the true density peaks and their relative distributions
with respect to the extinction.

Nevertheless, the majority of these cores appear to be bona fide dense
cores. We find that the fraction of dense cores within the Pipe
Nebula, as traced by \ammoniaoneone\, emission, is 100\% for all cores
with peak \av\, $>$ 15 magnitudes and mass $>$ 6\,\Msun. This drops to
$\sim$ 60\% for cores with masses $<$ 6\,\Msun. Deeper observations
and mapping could likely raise the detection rate for \ammonia\,
toward these cores.

\subsection{Measured velocity dispersions}

\begin{figure}
\begin{center}
\includegraphics[angle=90,width=0.45\textwidth]{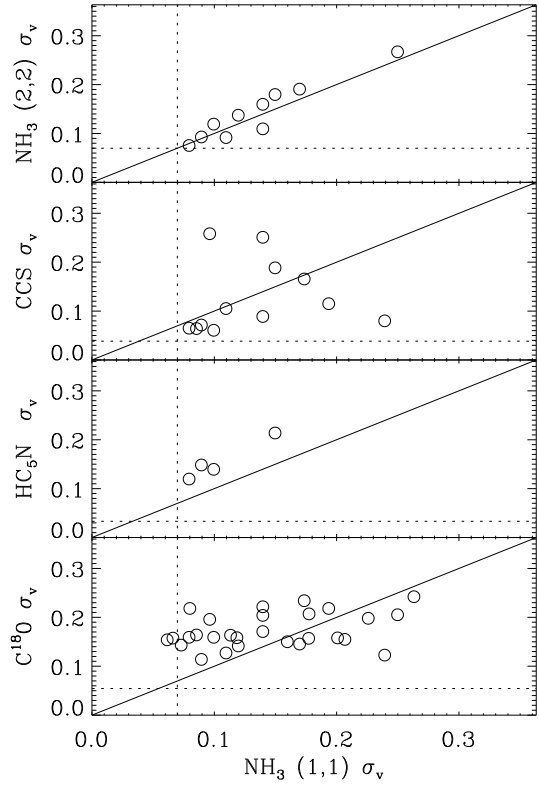}
\caption{\label{linewidths}Measured one dimensional velocity dispersion for cores with \ammoniatwotwo, CCS, 
\hcfivennt, and \ceont\, emission  compared to those measured from 
the \ammoniaoneone\, emission. All axes are in units of \kms. The
solid line traces equal values of the dispersions. The dotted lines
trace the thermal velocity dispersion for each molecule (assuming 10 K gas).}
\end{center}
\end{figure}

Figure~\ref{linewidths} shows the measured one dimensional velocity
dispersions for cores detected in \ammoniatwotwo, CCS, \hcfivennt, and
\ceont\, as a function of the velocity dispersion measured from the
\ammoniaoneone\, emission (the \ceont\, data are from \citealp{Gus-pipe}). For
gas with \tk = 10~K the thermal velocity dispersions for these
molecules are listed in Table~\ref{line-info} and are marked as dotted
lines in Figure~\ref{linewidths}.

These data reveal a strong correlation between the \ammoniatwotwo, \hcfivennt,
\ceont, and \ammoniaoneone\, velocity dispersions. In the case of
\ammoniatwotwo, the velocity dispersions are essentially 
identical to those measured from the \ammoniaoneone\, emission. For
the CCS to \ammoniaoneone\, comparison, we see a larger range in the
velocity dispersion. We find for \hcfivennt\, the velocity
dispersions are typically broader than those measured from the
\ammoniaoneone\, emission. Because \hcfivennt\, is heavier than 
the \ammonia\, molecule (and thus has a smaller thermal velocity
dispersion) the fact that the measured velocity dispersion from
\hcfivennt\, are typically broader than \ammonia, may indicate that
this emission is tracing a larger non-thermal component. With only
four detections it is difficult to determine this with certainty.

We find the velocity dispersions measured from the \ammonia\, emission
are typically narrower than those measured from the \ceont: the \ceont\,
velocity dispersions are typically 1.4 times broader than \ammonia.
This is expected given the \ceont\, emission will trace all the gas
along the line of sight, in contrast to \ammonia, which traces only
the gas associated with the dense compact core. As a result, the
\ceont\,  observations will over estimate the non-thermal
component of the velocity dispersion.

Most cores have velocity dispersions which are about twice their
thermal velocity dispersions; the exception being those traced by
\hcfivennt\, and \ceont\, which show that \sigmav\, is $\sim$ 3 to 4 times
broader than thermal.  This suggests that the \ammonia, CCS, \hcfivennt, and
\ceont\, emission is primarily tracing the non-thermal component of the
gas. We find that the measured velocity dispersions from the \ceont\,
and \hcfivennt\, are the comparable for those cores where emission was
detected in both transitions.  Because carbon species are more
depleted in the densest inner core regions, it is likely that both these
species arise in the outer most core edges.

We find three cores that have velocity dispersions comparable to the
thermal \ammonia\, dispersion (assuming gas at 10 K). Although these
cores have low signal to noise detections, it may be that they are
purely dominated by thermal processes. Using the unique measurement of
\tk\, for each of the 12 cores where the \ammoniaoneone\, and (2,2)
lines were sufficiently strong that the spectra could be modelled (the
cores listed in Table~\ref{fitspec}), we find that the majority of
these cores have measured velocity dispersions $\sim$ 1 to 2 times
broader than thermal. 

\begin{figure}
\begin{center}
\includegraphics[width=0.6\textwidth]{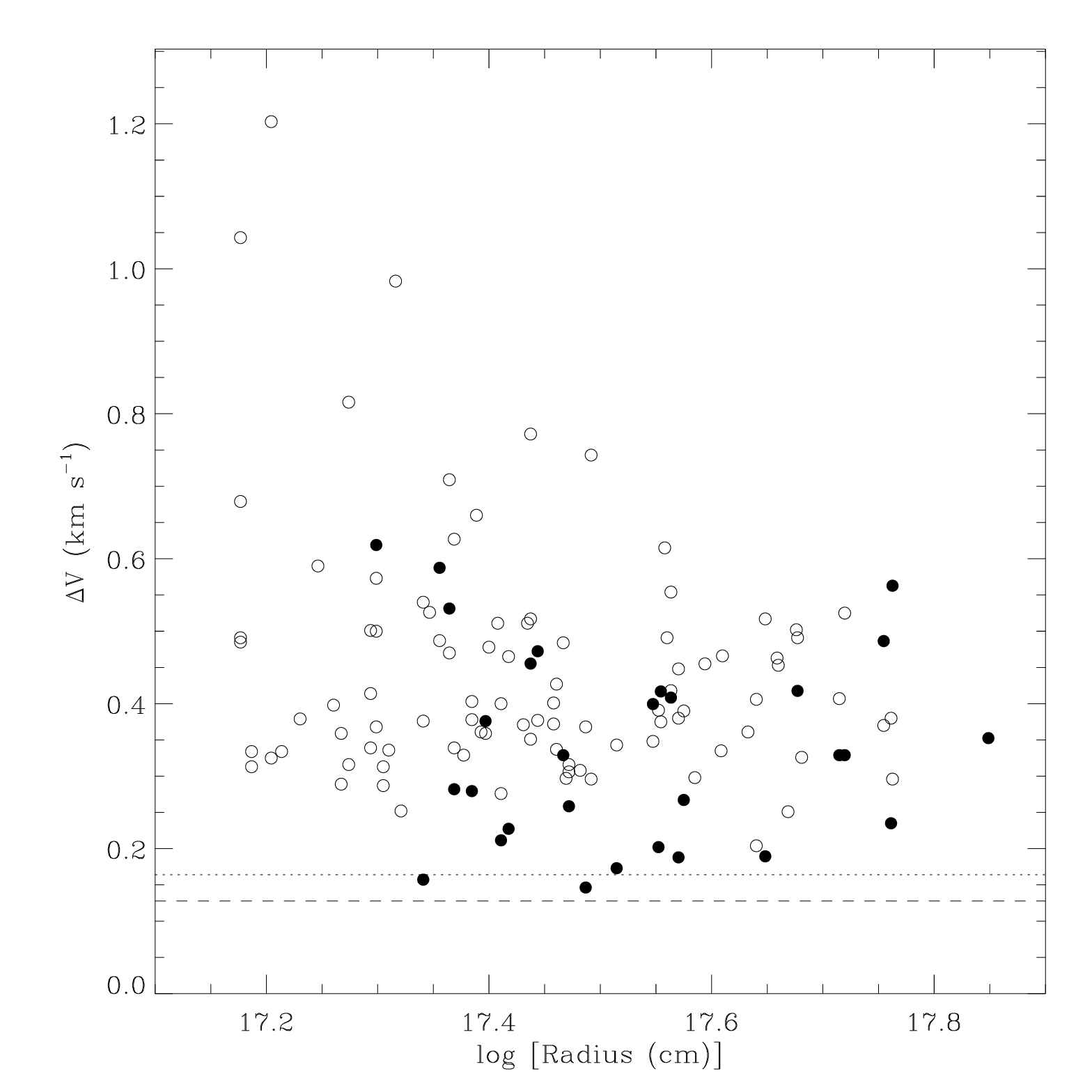}
\caption{\label{size-linewidth}Core radius versus linewidth for all cores detected 
in \ceont\, (open circles) and \ammonia\,(closed circles)
emission. While the linewidths measured from the \ceont\, are
typically broader than those measured from \ammonia\, emission, there
is no correlation between the radius and linewidth in either species
for the cores within the Pipe Nebula. The dotted and dashed lines
marks the thermal linewidths for \ammonia\,and \ceont\,
respectively.}
\end{center}
\end{figure}

With this large sample of cores we can also investigate the
size-linewidth relation (e.g. \citealp{Larson81}). The data presented
here have the advantage of arising from cores within the same cloud
which makes comparisons between these properties independent of
distance effects. Figure~\ref{size-linewidth} shows the radius versus
linewidth plot for cores detected in both \ceont\, and
\ammoniaoneone. While we see the \ceont\, cores (marked with open
circles, from \citealp{Gus-pipe}) typically have broader linewidths
compared to the \ammonia\, cores (marked as filled circles), the
general trend between these two samples is similar: there appears to
be no correlation between the core linewidth and size. As will be
discussed in more detail in \cite{Lada-pipe}, the relatively flat
distribution of the data within Figure~\ref{size-linewidth} suggests,
therefore, that the cores are dominated by thermal rather than
turbulent motions. This is also seen toward dense cores in other star
forming regions (e.g. \citealp{Barranco98,Goodman98}).

\subsection{Comparison to other star-forming regions}

Past \ammonia\, surveys of star forming regions such as Taurus,
Ophiuchus, Perseus, Orion, and Cepheus, reveal that the measured
properties within these regions showed general trends based on their
star formation activity (see \citealp{Jijina99} for a review).  For
instance, the regions dominated by dense young clusters (e.g. Orion
and Cepheus) tend to have higher measured \tk, linewidths, and core
masses. In contrast, Taurus and Ophiuchus have lower values for these
properties, however, these cores tend to have higher column
densities. Perseus, being intermediate in its star-formation activity,
also shows intermediate values for the derived \ammonia\, properties
(see \citealp{Jijina99} and referenecs therein).

Many of these studies find strong \ammonia\, emission and a high
detection rate toward the dense cores. However, many of the \ammonia\,
cores within these surveys are associated with \iras\, sources; in the
case of Orion, \cite{Benson89} found \iras\, sources coincident with
68\% of the \ammonia\, cores. Such a high fraction of cores with
obvious infrared emission suggests that star formation is already
occurring within these cores. Thus, the detection of strong \ammonia\,
emission in a starless core may imply that the star formation process
is imminent. This is consistent with chemical evolutionary models that
predict \ammonia\, to be most abundant in the later evolutionary
stages of core formation \citep{Suzuki92}.

Our lower \ammonia\, detection rate, compared to other \ammonia\,
surveys of star-forming regions, is consistent with the fact that the
Pipe cores have lower densities and most show little signs of active
star formation. Indeed, the core associated with B59, the only site
within the Pipe Nebula where current star formation is known to be
occurring, is one of the few cores that shows very strong \ammonia\,
emission. Although the detected \ammonia\, emission was typically weak
toward our sample of cores, these cores do harbor the densest gas
within the Pipe Nebula. Moreover, those cores that show strong CCS
emission may be the very youngest within the region.

\begin{figure}
\begin{center}
\includegraphics[angle=90,width=0.45\textwidth]{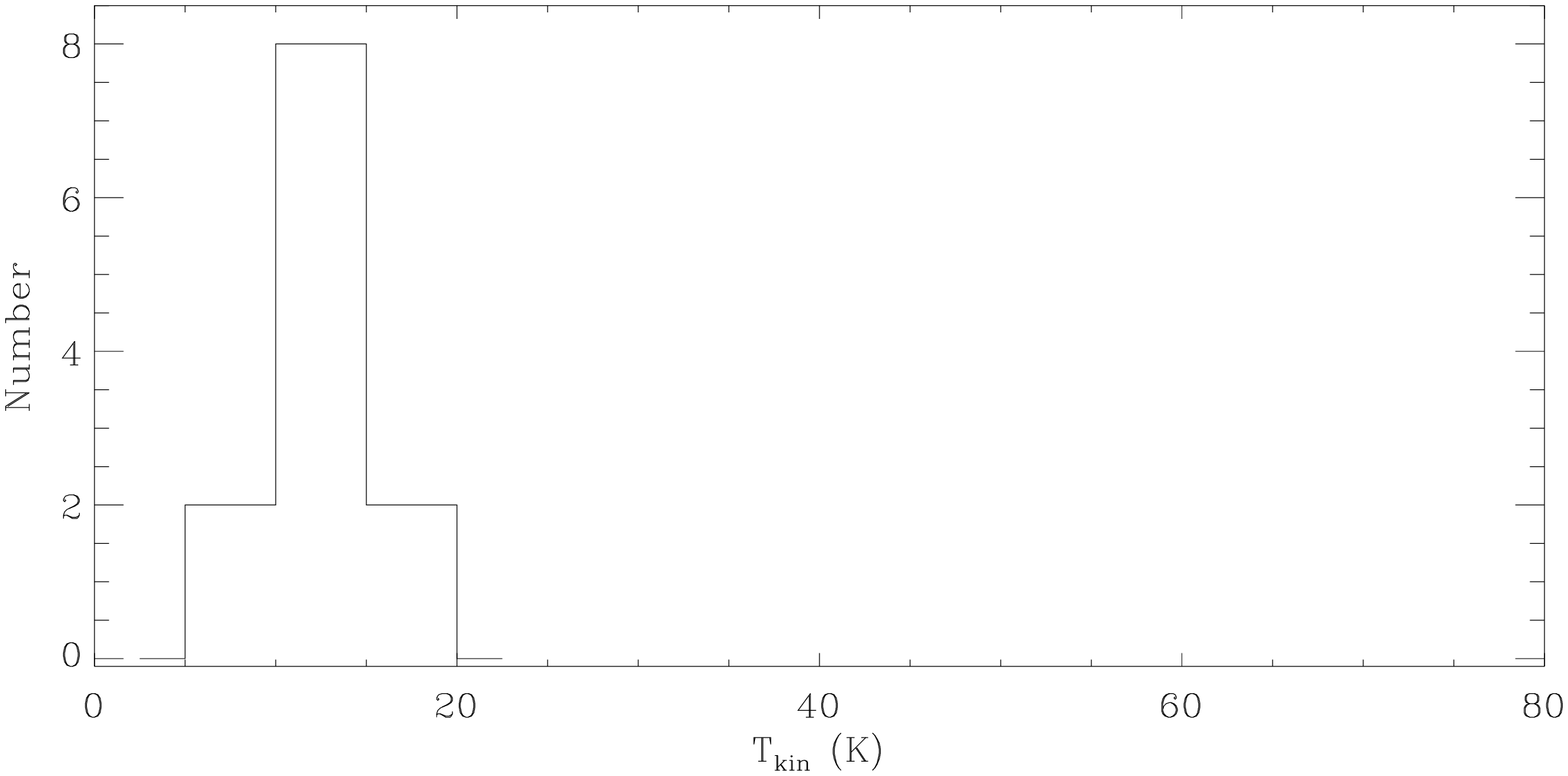}\\
\includegraphics[angle=90,width=0.45\textwidth]{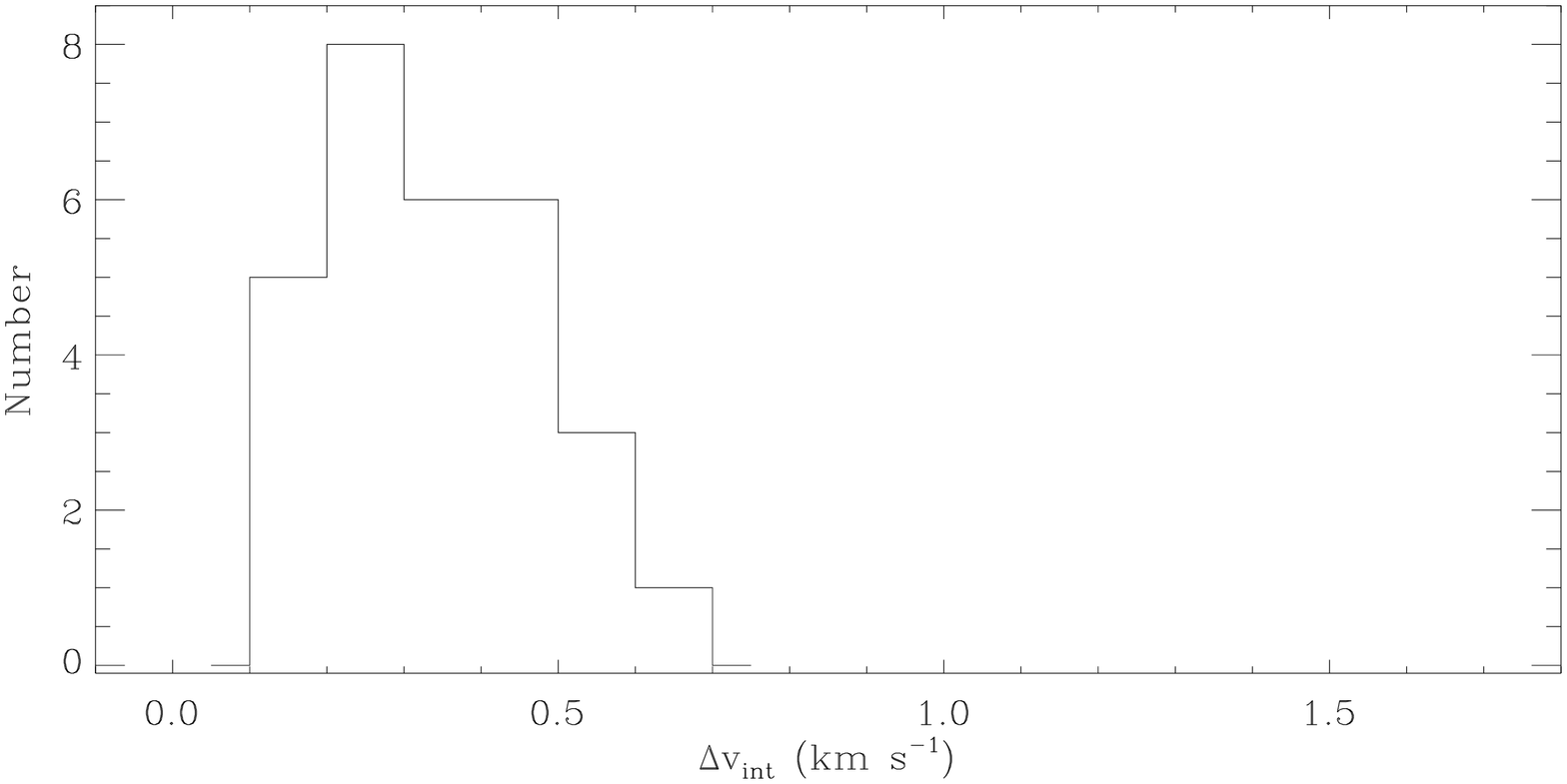}\\
\includegraphics[angle=90,width=0.45\textwidth]{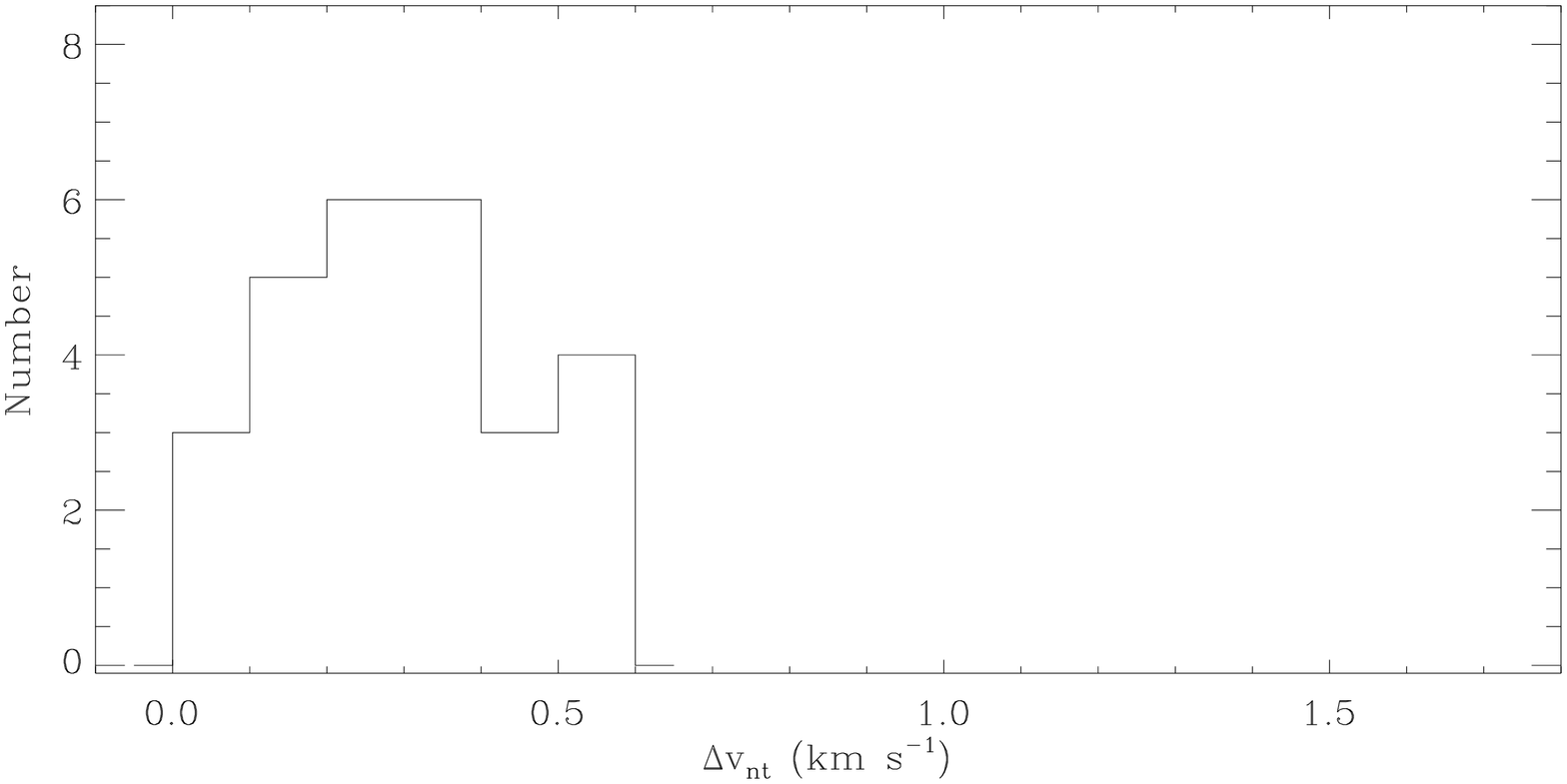}\\
\includegraphics[angle=90,width=0.45\textwidth]{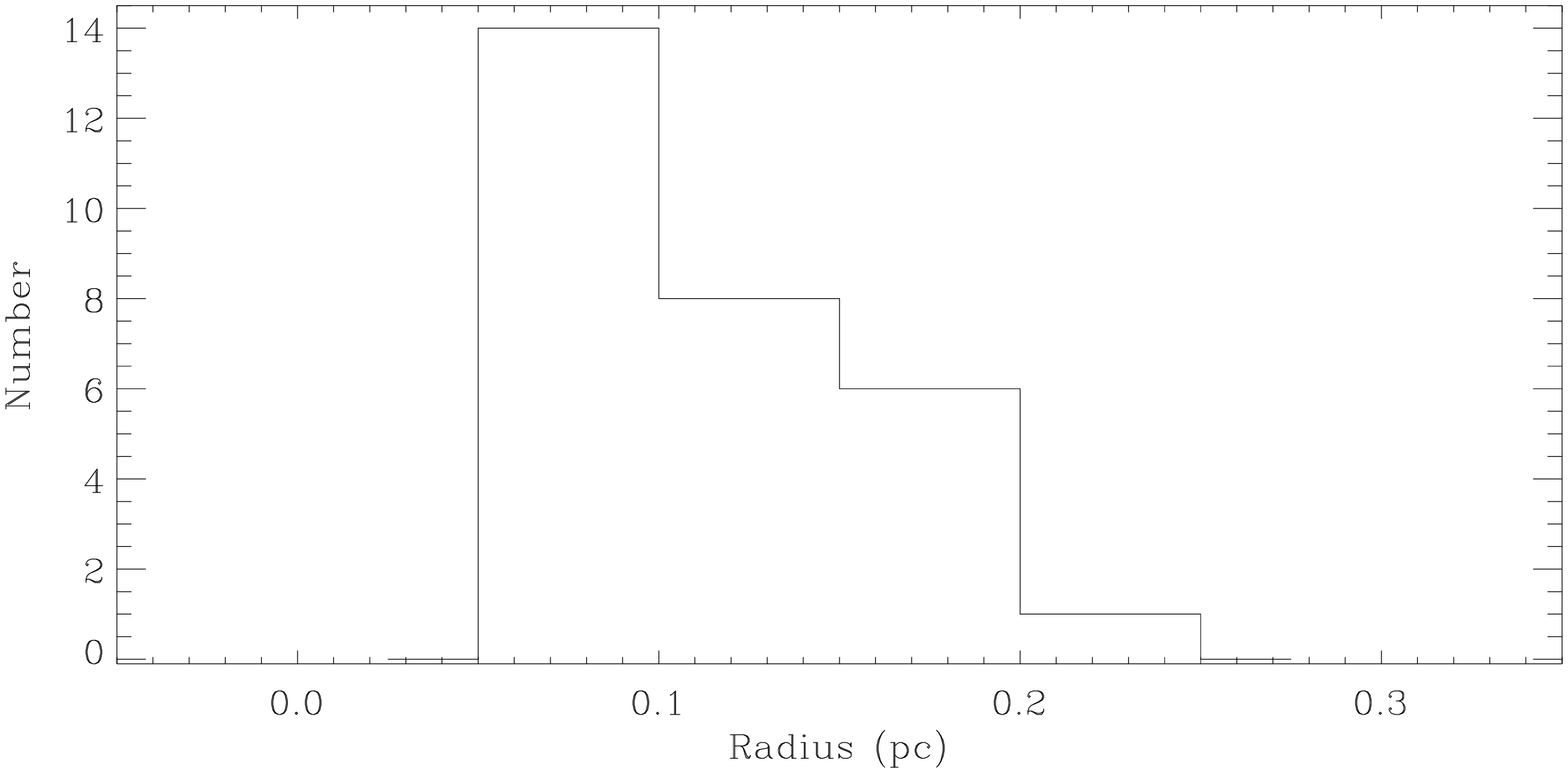}\\
\includegraphics[angle=90,width=0.45\textwidth]{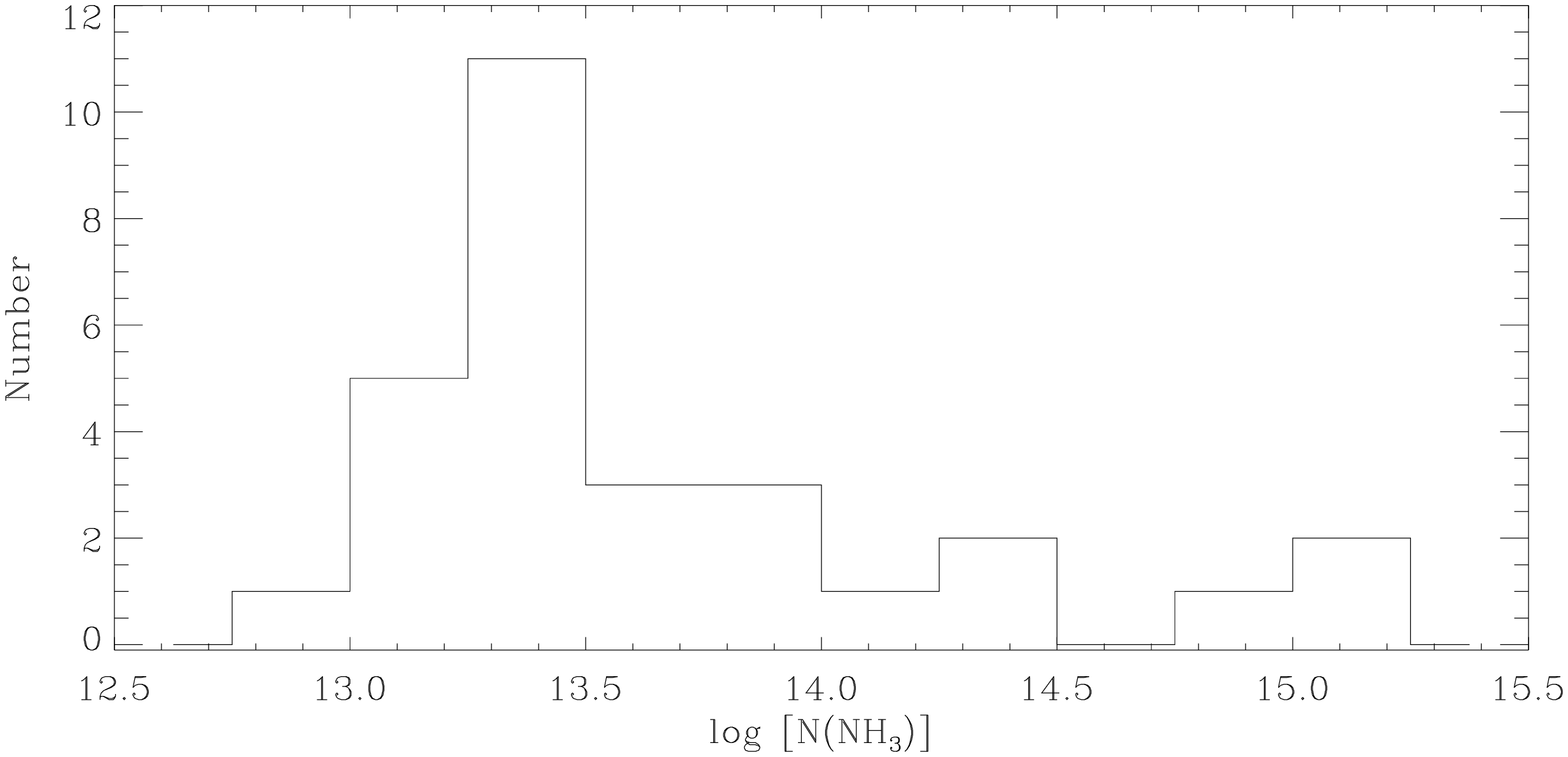}
\caption{\label{property-comparisons}Histograms of the derived properties for the 
cores within the Pipe Nebula. Comparisions between these histograms
and similar ones for other star forming regions (e.g. Taurus,
Ophiuchus, Perseus, Orion, and Cepheus) reveal the Pipe cores are most
similar to cores within Perseus and Opiuchus. The obvious exception is that the
Pipe cores have lower column densities and that the region contains
very little active star formation.}
\end{center}
\end{figure}

Comparing our measured values for \tk, the intrinsic $\Delta$V, the
non-thermal $\Delta$V, radius, and column density
(Fig.~\ref{property-comparisons}) to those measured in other star
forming regions reveals that the cores in the Pipe Nebula are most
similar to cores within Perseus (see figures within
\citealp{Jijina99}). The obvious difference between the Perseus cores 
and those within the Pipe Nebula is the derived column densities and
the star formation activity: the Pipe cores show lower column
densities and very little star formation.

The lower \ammonia\, column densities derived for the Pipe cores could arise
because of differences in the intrinsic extinction,  abundance of
\ammonia, or excitation conditions compared to other star-forming regions.
Because the cores within the Pipe Nebula span a range in peak
extinction and core mass similar to other star-forming regions, it
seems unlikey that the lower measured column densities arise simply
due to differences in their measured extinction. Moreover, the
measured column densities of the cores show an order of magnitude
greater range (10$^{13}$--10$^{15}$\,\cms) compared to the peak
extinction (5--20\,mags).  A more likely scenario is that the
intrinsic abundance of \ammonia\, is low within many of these cores,
possibly due to their relative chemical youth. It may be, however,
that the excitation conditions with these cores are insufficient to
significantly populate the levels. It is possible that our derived
\ammonia\, column densities are under estimates and that \ammonia\, is more
abundant within the cores, but simply not excited to the levels
required for strong detections. This is consistent with the relatively
low mean densities of the cores ($\sim$ 7 $\times$ 10$^{3}$\,\cmc)
compared to that necessary to sufficiently populate the observed
levels ($\sim$ 10$^{4}$\,\cmc).  Detailed radiative transfer modelling
is required to distinguish between these scenarios.

If we restrict our sample of cores to those with similar column
densities as measured toward other star-forming regions
(i.e. N(\ammonia) $>$ 10$^{14}$\,\cmc), we find that the cores within
the Pipe share similar properties with cores in both Perseus and
Ophiuchus. Given that only 6 cores were detected in the Pipe with such
high column densities, it is difficult to determine the distribution
in their properties with any certainty. Nevertheless, it appears that
the cores in the Pipe Nebula are similar to those found in other
low-mass star forming regions. The most noticable difference is that
the majority of Pipe cores tend to show weaker \ammonia\, emission and
have little star-formation activity.

\subsection{Core evolution as traced by column density}

The chemical characteristics within dense cold cores can reveal their
evolutionary stage. Because they have different production
chemistries, the column densities of \ammonia, CCS,
\hcfivennt, and \ceont\, can reveal the chemical characteristics in
dense cores (e.g. \citealp{Suzuki92}).  For instance, models for core
evolution \citep{Suzuki92} predict an anti-correlation between the
abundance of CCS and \ammonia; with CCS more abundant in the early
evolutionary phases.

Because of the evolutionary effects of depletion and the differences
in the molecular distributions within the cores, a single measurement
of the column density within the core can only provide limited
information on the chemical structure and evolutionary stage.
Nevertheless, we find that the comparisons between the column
densities of the various species observed within the Pipe cores are
similar to what is seen toward dense cores in Taurus and Ophiuchus
\citep{Benson89,Suzuki92}.  In particular, we find that: (1) the
N(\ammonia) and N(CCS) show no clear correlation, which is consistent
with differences in the densities of the cores
(0.5--2.5$\times$10$^{4}$\,\cmc); (2) the N(\ceont) is essentially
constant for all values of the N(CCS), which implies that the CCS is a
more reliable tracer of denser gas; and (3) that N(CCS) and
N(\hcfivennt) are correlated, which is expected given that these
carbon-chain molecules are produced under the same chemical conditions
within a core \citep{Fuente90,Suzuki92}.

We also find high abundance ratios of [\ammonia] compared to [\ceont]
for the densest cores, confirming that \ammonia\, is a better tracer
of dense gas than \ceont\, and that \ceont\, traces the outer core
envelope and is depleted in the core centers.  Because the chemical
abundances trace the evolutionary age of the cores, we can use the
fractional abundances of these molecules to estimate the evolutionary
ages of the cores. Using the model calculations of \cite{Suzuki92},
we estimate the cores are at an evolutionary age of $\sim$ 6 $\times$
10$^{5}$ years.

%%%%%%%%%%%%%%%%%%%%%%%%%%%%%%%%%%%%%%%%%%%%%%%%%%%%%%%%%%%%%%%%%%%%%%%%%%%%%%%%%%%%
\section{Conclusions}

Using the 100 m GBT we have conducted an \ammoniaoneone,
\ammoniatwotwo, \ccs, and \hcfiven\, survey toward a sample of 46
cores within the Pipe Nebula. These cores were identified via their
high extinction and appear to be dense starless cores which likely
represent the initial conditions of star formation. In total, we
detect \ammoniaoneone\, emission in 29 cores ($\sim$~63\%), \ammoniatwotwo\,
emission in 10 cores ($\sim$~22\%), CCS emission in 13 cores
($\sim$~28\%), and \hcfivennt\, emission in 4 cores ($\sim$~9\%). Not
surprisingly, the detection rate is higher toward those cores with
higher peak \av\, and higher mass; in the case of \ammoniaoneone\, we
detected emission toward all cores with a peak \av $>$ 15 magnitudes
and with masses~$>$~11\,\Msun.

We find that the cores are associated with dense ($>$10$^{4}$\,\cmc)
gas. The \ammonia\, emission revealed n(\hh) typically three times
greater than the mean densities measured from the extinction
maps. This is expected given that the cores likely have inwardly
increasing density gradients; the \ammonia\, is tracing the densest
centers of the cores.

Using a forward-fitting modelling routine, we estimated the physical
properties of the gas (\tk, $\tau$, and \tex) for 12 cores. We find
that \tk\, ranges from 9.5 to 17 K, with a median value of $\sim$
12 $\pm$ 2~K. Cores with higher mass tend to have lower \tk\, values
(10 $\pm$ 1 K) compared to lower-mass cores (13 $\pm$ 3 K).

The measured one dimensional velocity dispersion, \sigmav, for the
cores show that the lines are narrow, with several cores having
dispersions close to thermal (for a 10 K gas). A comparison between
the linewidths measured for the cores from \ceont\, and \ammonia,
reveal that while the general trends are the same, i.e.\, neither
dataset show a size-linewidth relation, the \ammonia\, linewidths
are systematically lower than those measured from the \ceont\,
emission.  This arises because the \ceont\, traces all emission along
the line of sight, whereas \ammonia\, traces only the densest gas
associated with the core.

The derived properties of the Pipe cores are similar to dense cores
within other star-forming regions, i.e.\,Perseus, Taurus,
Ophiuchus. The most obvious difference is that the cores within the
Pipe Nebula typically have weaker \ammonia\, emission and most show no
evidence for current star formation. The weak \ammonia\, emission may arise due
to the low abundance of \ammonia\, because the cores are chemically
young or because the excitation conditions (densities) within the gas
are insufficient to excite the observed transitions.  Detailed
radiative transfer models are needed to resolve this issue.

Although we have a limited number of cores, we estimate their
evolutionary ages to be $\sim$ 6 $\times$ 10$^{5}$ years.  Thus, the
cores within the Pipe Nebula represent an excellent sample of dense
cores in which to study the initial conditions of star-formation and
the earliest stages of core formation and evolution.

%%%%%%%%%%%%%%%%%%%%%%%%%%%%%%%%%%%%%%%%%%%%%%%%%%%%%%%%%%%%%%%%%%%%%%%%%%%%%%%%%%%%
\acknowledgments

We are extremely grateful to Erik Rosolowsky for the use of the
\ammonia\, forward-fitting routine ahead of publication and for the many 
informative discussions.  We also thank the support staff at the GBT,
in particular Frank Ghigo, for help will the observing setup and data
reduction.  We acknowledge funding support through NASA Origins grant
NAG-13041.

\appendix
\section{Appendix}

\begin{figure}[ht]
\includegraphics[angle=90,width=0.9\textwidth]{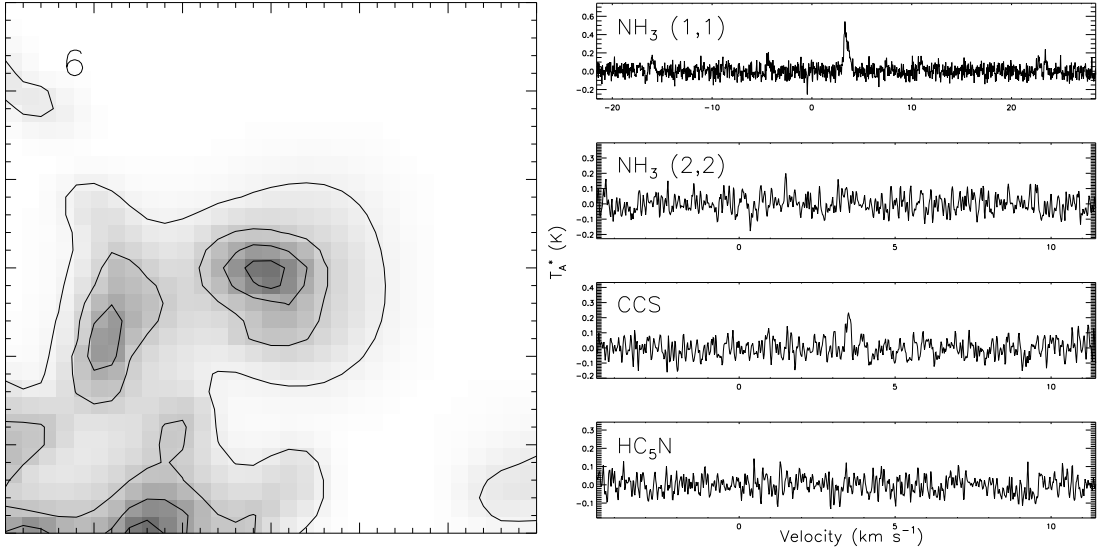}
\caption{\label{appendix-fig-12a} The extinction image (left) and spectra (right) toward core 6. See the appendix text for
a description of the image and spectra.}
\end{figure}

The 2MASS extinction map and \ammoniaoneone, \ammoniatwotwo, \ccs, and \hcfiven\,
spectra for each core are shown in Figs~\ref{appendix-fig-12a} to
55. Labelled in the top left corner of each
extinction image is the core number based on the catalog of
\cite{Alves07}. In all cases, the extinction images are 0\fdg25
$\times$ 0\fdg25 and are centered on the peak extinction for the core
(as determined from these maps). The grey scale is from an \av\, of 0
to 20 magnitudes. The contour levels are 1.2, 4, 6.8, 9.6, and 12.4
magnitudes (3$\sigma$ in steps of 7$\sigma$). The final image in the
appendix, Figure~56, summarises the details of the
extinction images and includes the sizes of the 2MASS extinction image
angular resolution and the GBT beam.

Each spectrum is centered on the \vlsr\, of the core (see
Tables~\ref{line-parameters-1} and \ref{line-parameters-2}). The
\ammoniaoneone\, spectra extend $\pm$ 25 \kms\, while the remaining 
three spectra extend $\pm$ 8 \,\kms\, around the core's \vlsr. The
temperature scale is \tastar.

%%%%%%%%%%%%%%%%%%%%%%%%%%%%%%%%%%%%%%%%%%%%%%%%%%%%%%%%%%%%%%%%%%%%%%%%%%%%%%%%%%%%
% References

%%%%%%%%%%%%%%%%%%%%%%%%%%%%%%%%%%%%%%%%%%%%%%%%%%%%%%%%%%%%%%%%%%%%%%%%%%%%%%%%%%%%
% Tables
%%%%%%%%%%%%%%%%%%%%%%%%%%%%%%%%%%%%%%%%%%%%%%%%%%%%%%%%%%%%%%%%%%%%%%%%%%%%%%%%%%%%

%%%%%%%%%%%%%%%%%%%%%%%%%%%%%%%%%%%%%%%%%%%%%%%%%%%%%%%%%%%%%%%%%%%%%%%%%%%%%%%%%%%%

\begin{table}
\begin{center}
\caption{\label{line-info}Molecular transition frequencies and constants.}
\begin{tabular}{llccccc}
\tableline  \tableline
\multicolumn{1}{c}{Molecule}  & \multicolumn{1}{c}{Transition}  & $\nu$  & $\mu$ & E$_{u}$ & B & Thermal \sigmav \tablenotemark{a} \\
          &             & (GHz)  & (D)   & (cm$^{-1}$) & (GHz) & (\kms)\\
\tableline
\ammonia\, & (J,K) = (1,1)              &  23.694 & 1.47  & 16.25 & 298.117  & 0.07\\
\ammonia\, & (J,K) = (2,2)              &  22.733 & 1.47  & 45.08 & 298.117  & 0.07 \\
CCS        & J$_{N}$ = 2$_{1}$--1$_{0}$ &  22.344 & 2.81  & 1.12  & 6.478    & 0.04\\
\hcfivennt & J = 9--8                   &  23.963 & 4.33  & 4.00  & 1.331    & 0.03\\
\ceont\,\tablenotemark{b}   & J = 1--0  & 109.782 & 0.11  & 3.66 & 54.890  & 0.05 \\
\tableline
\end{tabular}
\tablenotetext{a}{Assuming a gas temperature of 10 K.}
\tablenotetext{b}{These data are from \cite{Gus-pipe}. The parameters are included here for completeness
when discussing the calculation of column density.}
%\tablerefs{need to add in references here.....}
\end{center}
\end{table}

%%%%%%%%%%%%%%%%%%%%%%%%%%%%%%%%%%%%%%%%%%%%%%%%%%%%%%%%%%%%%%%%%%%%%%%%%%%%%%%%%%%%

\begin{table}
{\scriptsize{
\begin{center}
\caption{\label{table1}Observing parameters}
\begin{tabular}{rccccccccccc}
\tableline  \tableline
Core\tablenotemark{a} & \multicolumn{2}{c}{Coordinates} & Peak   & Mass    & log[n(\hh)]\tablenotemark{b}   &  Date & Int. & El & \tsys & 1$\sigma$ rms   \\
     & RA      & Dec                   & \av      &         &             &       & time &    &       &                 \\
     & (J2000) & (J2000)               & (mag) & (\Msun) & (cm$^{-3}$) &       & (mins)     & (\degree) & (K)   & (K) \\     
\tableline
  6*   &  17:10:31.57  &  $-$27:25:51.59  &  13.6  &     3.1  &     3.9  &  2006-10-18  &       22  &       22  &   96  &   0.04   \\      %Pipe-12a
  7*   &  17:11:36.95  &  $-$27:33:27.08  &  12.1  &     4.7  &     3.8  &  2006-11-05  &       18  &       17  &   77  &   0.04   \\      %Pipe-17a
  8*   &  17:12:12.59  &  $-$27:37:20.94  &  12.2  &     3.3  &     3.9  &  2006-10-18  &       18  &       19  &  102  &   0.06   \\      %Pipe-15a
 11*   &  17:10:51.00  &  $-$27:22:59.54  &  11.5  &     3.4  &     3.8  &  2006-11-05  &       18  &       22  &   69  &   0.03   \\      %Pipe-22a
 12*   &  17:11:20.49  &  $-$27:26:28.98  &  18.0  &    20.4  &     3.8  &  2006-10-13  &        5  &       16  &   60  &   0.06   \\      %Pipe-B59c-2
 13\,\,&  17:10:49.69  &  $-$27:13:25.01  &   5.3  &     0.5  &     3.9  &  2006-10-29  &       18  &       19  &   77  &   0.04   \\      %Pipe-131
 14*   &  17:12:34.03  &  $-$27:21:16.24  &  18.5  &     9.7  &     3.9  &  2006-10-13  &       18  &       23  &   52  &   0.02   \\      %Pipe-5
 15*   &  17:12:53.38  &  $-$27:23:23.98  &  10.4  &     2.6  &     3.8  &  2006-11-05  &       18  &       23  &   70  &   0.03   \\      %Pipe-33a
 17\,\,&  17:14:06.81  &  $-$27:28:25.97  &   5.3  &     0.7  &     3.9  &  2006-10-29  &       69  &       20  &   48  &   0.01   \\      %Pipe-129
 20*   &  17:15:11.16  &  $-$27:35:06.00  &   9.6  &     2.3  &     3.8  &  2006-11-05  &       18  &       21  &   74  &   0.04   \\      %Pipe-44a
 22*   &  17:15:48.00  &  $-$27:29:32.86  &   7.7  &     1.0  &     3.9  &  2007-02-08  &       18  &       16  &   50  &   0.02   \\      %Pipe-73a
 23*   &  17:16:07.19  &  $-$27:31:11.97  &   8.4  &     1.9  &     3.8  &  2006-10-25  &       74  &       19  &   75  &   0.02   \\      %Pipe-58a
 25*   &  17:16:23.68  &  $-$27:10:11.98  &   6.6  &     1.1  &     3.8  &  2007-02-08  &       18  &       18  &   47  &   0.02   \\      %Pipe-104a
 27*   &  17:17:07.04  &  $-$27:01:48.05  &   6.6  &     3.1  &     3.7  &  2006-11-05  &       18  &       24  &   68  &   0.03   \\      %Pipe-102a
 30\,\,&  17:21:01.87  &  $-$27:13:42.65  &   5.3  &     0.4  &     3.8  &  2006-10-29  &       18  &       23  &   68  &   0.03   \\      %Pipe-128
 31*   &  17:18:32.07  &  $-$26:49:33.57  &   6.6  &     1.9  &     3.7  &  2006-11-05  &       18  &       21  &   77  &   0.04   \\      %Pipe-101a
 33*   &  17:19:35.00  &  $-$26:55:47.96  &  10.7  &     4.3  &     3.8  &  2006-11-05  &       18  &       21  &   72  &   0.04   \\      %Pipe-31a
 34*   &  17:20:18.86  &  $-$26:59:18.63  &   8.4  &     2.7  &     3.8  &  2006-11-05  &       18  &       24  &   68  &   0.03   \\      %Pipe-61a
 36\,\,&  17:19:29.25  &  $-$26:46:18.90  &   6.6  &     1.7  &     3.8  &  2006-10-29  &       18  &       24  &   66  &   0.03   \\      %Pipe-105
 37*   &  17:19:32.21  &  $-$26:43:30.06  &   8.8  &     2.0  &     4.1  &  2006-10-24  &       18  &       16  &   76  &   0.04   \\      %Pipe-56a
 40*   &  17:21:16.43  &  $-$26:52:56.68  &  23.7  &     9.2  &     3.8  &  2006-10-13  &        5  &       24  &   50  &   0.05   \\      %Pipe-1
 41*   &  17:22:29.10  &  $-$27:04:03.15  &   9.8  &     1.1  &     4.0  &  2006-11-26  &       18  &       20  &   75  &   0.04   \\      %Pipe-40a
 42*   &  17:22:42.12  &  $-$27:05:00.58  &  21.6  &     2.8  &     4.1  &  2006-10-13  &       18  &       24  &   50  &   0.02   \\      %Pipe-3
 47*   &  17:27:29.55  &  $-$26:59:05.97  &  10.3  &     1.4  &     3.8  &  2006-10-24  &       18  &       21  &   66  &   0.03   \\      %Pipe-34a
 48*   &  17:25:59.04  &  $-$26:44:11.78  &  11.1  &     4.2  &     3.7  &  2006-10-18  &       18  &       15  &  120  &   0.07   \\      %Pipe-27a
 51*   &  17:27:24.00  &  $-$26:44:23.99  &   8.6  &     1.2  &     3.8  &  2006-10-24  &       18  &       14  &   90  &   0.04   \\      %Pipe-57a
 56*   &  17:28:10.86  &  $-$26:24:01.16  &  11.5  &     5.2  &     3.8  &  2006-10-18  &       18  &       18  &  107  &   0.06   \\      %Pipe-21a
 61*   &  17:28:38.31  &  $-$26:16:55.20  &   8.9  &     2.6  &     3.8  &  2006-11-26  &       18  &       17  &   84  &   0.04   \\      %Pipe-52a
 62*   &  17:28:47.94  &  $-$26:18:27.05  &   9.7  &     2.3  &     3.7  &  2006-11-05  &       18  &       24  &   68  &   0.03   \\      %Pipe-42a
 65*   &  17:31:20.65  &  $-$26:30:36.05  &  11.6  &     0.7  &     4.1  &  2006-10-18  &       18  &       20  &   98  &   0.05   \\      %Pipe-20a
 66*   &  17:31:15.79  &  $-$26:29:06.03  &  10.9  &     1.0  &     4.0  &  2006-11-05  &       18  &       19  &   84  &   0.04   \\      %Pipe-29a
 70*   &  17:29:35.59  &  $-$25:54:23.39  &  11.3  &     1.1  &     4.0  &  2006-11-05  &       18  &       17  &   91  &   0.05   \\      %Pipe-24a
 74*   &  17:32:35.28  &  $-$26:15:54.01  &  12.4  &     3.0  &     3.7  &  2006-11-05  &       18  &       19  &   75  &   0.04   \\      %Pipe-14a
 87*   &  17:34:11.47  &  $-$25:50:15.01  &  21.0  &    10.3  &     3.9  &  2006-10-14  &        5  &       13  &   67  &   0.07   \\      %Pipe-2a
 89*   &  17:33:26.56  &  $-$25:40:12.01  &  13.7  &     1.4  &     4.0  &  2006-11-05  &       74  &       25  &   70  &   0.02   \\      %Pipe-11a
 91*   &  17:32:15.63  &  $-$25:25:02.97  &  11.4  &     1.1  &     4.1  &  2006-11-05  &       18  &       15  &   99  &   0.05   \\      %Pipe-23a
 92*   &  17:34:06.15  &  $-$25:40:03.02  &  15.1  &     1.6  &     4.0  &  2006-10-18  &       18  &       25  &   87  &   0.04   \\      %Pipe-8a
 93*   &  17:34:45.21  &  $-$25:46:57.09  &  16.9  &     3.5  &     3.9  &  2007-02-08  &       18  &       23  &   43  &   0.02   \\      %Pipe-7a
 97*   &  17:33:30.19  &  $-$25:31:11.97  &  14.6  &     5.9  &     3.6  &  2006-11-05  &       18  &       14  &   83  &   0.04   \\      %Pipe-9a
 99*   &  17:25:06.50  &  $-$24:12:48.37  &   7.3  &     2.2  &     4.0  &  2006-11-05  &       18  &       22  &   76  &   0.04   \\      %Pipe-85a
101*   &  17:22:43.12  &  $-$23:50:08.51  &  12.8  &     1.9  &     4.1  &  2006-11-26  &       18  &       21  &   74  &   0.04   \\      %Pipe-13a
102*   &  17:34:17.10  &  $-$25:34:12.02  &  19.2  &     6.7  &     3.6  &  2007-02-08  &       18  &       25  &   43  &   0.02   \\      %Pipe-4a
108\,\,&  17:31:34.13  &  $-$24:58:55.22  &   7.4  &     0.8  &     3.8  &  2006-10-25  &       18  &       19  &   68  &   0.03   \\      %Pipe-83
109*   &  17:35:48.49  &  $-$25:33:05.76  &  18.3  &     3.6  &     3.9  &  2006-10-13  &        5  &       17  &   58  &   0.07   \\      %Pipe-6
113*   &  17:23:35.71  &  $-$23:41:05.63  &   9.2  &     2.4  &     4.0  &  2006-10-24  &       18  &       21  &   68  &   0.03   \\      %Pipe-49a
132\,\,&  17:37:52.80  &  $-$25:14:57.51  &  11.2  &     4.7  &     3.7  &  2006-11-26  &       18  &       17  &   83  &   0.05   \\      %Pipe-26
\tableline
\end{tabular}
\tablenotetext{a}{Asterisks denote cores for which the position and value of the peak \av\, was determined from the higher-angular resolution images obtained with the NTT \citep{Carlos07}. The masses for these cores were, however, determined from the 2MASS background subtracted extinction maps.}
\tablenotetext{b}{Calculated from the visual extinction, assuming a spherical volume with radius equal to half the measured size of the core (as listed in \citealp{Alves07}).}
\end{center}}}
\end{table}

%%%%%%%%%%%%%%%%%%%%%%%%%%%%%%%%%%%%%%%%%%%%%%%%%%%%%%%%%%%%%%%%%%%%%%%%%%%%%%%%%%%%

\begin{table}
{\scriptsize{
\begin{center}
\caption{\label{line-parameters-1}Observed line parameters of \ammoniaoneone\, and \ammoniatwotwo.}
\begin{tabular}{rcccccccccc}
\tableline  \tableline
Core\tablenotemark{a} & & \multicolumn{4}{c}{\ammoniaoneone}    &  & \multicolumn{4}{c}{\ammoniatwotwo} \\
\cline{2-6} \cline{8-11}
     & & \tastar\, & \vlsr & \sigmav & I &  & \tastar\, & \vlsr & \sigmav & I  \\
     & & (K) & (\kms) & (\kms) & (\Kkms) & & (K)  & (\kms) & (\kms) & (\Kkms)  \\
\tableline
  6\,\,  &  &   0.41  &   3.43  &   0.09  &   0.08  &  &   --    &   --    &   --    &   --   \\      %Pipe-12a
  7\,\,  &  &   0.12  &   3.83  &   0.08  &   0.02  &  &   --    &   --    &   --    &   --   \\      %Pipe-17a
  8\,\,  &  &   0.17  &   3.47  &   0.11  &   0.05  &  &   --    &   --    &   --    &   --   \\      %Pipe-15a
 11\,\,  &  &   --    &   --    &   --    &   --  &    &   --    &   --    &   --    &   --   \\      %Pipe-22a
 12*     &  &   4.87  &   3.20  &   0.15  &   1.72  &  &   1.37  &   3.20  &   0.18  &   0.58   \\      %Pipe-B59c-2
 13\,\,  &  &   --    &   --    &   --    &   --    &  &   --    &   --    &   --    &   --   \\      %Pipe-131
 14*     &  &   0.42  &   3.47  &   0.14  &   0.14  &  &   0.09  &   3.52  &   0.11  &   0.02   \\      %Pipe-5
 15\,\,  &  &   0.17  &   3.60  &   0.18  &   0.07  &  &   --    &   --    &   --    &   --   \\      %Pipe-33a
 17*     &  &   0.20  &   3.38  &   0.25  &   0.12  &  &   0.06  &   3.41  &   0.27  &   0.04   \\      %Pipe-129
 20*     &  &   0.54  &   3.54  &   0.17  &   0.22  &  &   0.15  &   3.55  &   0.19  &   0.07   \\      %Pipe-44a
 22\,\,  &  &   0.12  &   3.72  &   0.12  &   0.03  &  &   --    &   --    &   --    &   --   \\      %Pipe-73a
 23\,\,  &  &   0.09  &   3.58  &   0.07  &   0.02  &  &   --    &   --    &   --    &   --   \\      %Pipe-58a
 25\,\,  &  &   0.13  &   3.65  &   0.20  &   0.06  &  &   --    &   --    &   --    &   --   \\      %Pipe-104a
 27\,\,  &  &   --    &   --    &   --    &   --    &  &   --    &   --    &   --    &   --   \\      %Pipe-102a
 30\,\,  &  &   --    &   --    &   --    &   --    &  &   --    &   --    &   --    &   --   \\      %Pipe-128
 31\,\,  &  &   --    &   --    &   --    &   --    &  &   --    &   --    &   --    &   --   \\      %Pipe-101a
 33\,\,  &  &   --    &   --    &   --    &   --    &  &   --    &   --    &   --    &   --   \\      %Pipe-31a
 34\,\,  &  &   --    &   --    &   --    &   --    &  &   --    &   --    &   --    &   --   \\      %Pipe-61a
 36\,\,  &  &   --    &   --    &   --    &   --    &  &   --    &   --    &   --    &   --   \\      %Pipe-105
 37\,\,  &  &   --    &   --    &   --    &   --    &  &   --    &   --    &   --    &   --   \\      %Pipe-56a
 40*     &  &   0.98  &   3.30  &   0.10  &   0.23  &  &   0.14  &   3.34  &   0.12  &   0.04   \\      %Pipe-1
 41*     &  &   0.39  &   3.69  &   0.12  &   0.11  &  &   0.11  &   3.72  &   0.14  &   0.04   \\      %Pipe-40a
 42*     &  &   0.43  &   3.75  &   0.11  &   0.11  &  &   0.10  &   3.87  &   0.09  &   0.02   \\      %Pipe-3
 47*     &  &   0.22  &   2.81  &   0.14  &   0.07  &  &   --    &   --    &   --    &   --   \\      %Pipe-34a
 48\,\,  &  &   --    &   --    &   --    &   --    &  &   --    &   --    &   --    &   --   \\      %Pipe-27a
 51\,\,  &  &   --    &   --    &   --    &   --    &  &   --    &   --    &   --    &   --   \\      %Pipe-57a
 56\,\,  &  &   --    &   --    &   --    &   --    &  &   --    &   --    &   --    &   --   \\      %Pipe-21a
 61\,\,  &  &   --    &   --    &   --    &   --    &  &   --    &   --    &   --    &   --   \\      %Pipe-52a
 62\,\,  &  &   --    &   --    &   --    &   --    &  &   --    &   --    &   --    &   --   \\      %Pipe-42a
 65\,\,  &  &   0.22  &   4.98  &   0.26  &   0.13  &  &   --    &   --    &   --    &   --   \\      %Pipe-20a
 66\,\,  &  &   --    &   --    &   --    &   --    &  &   --    &   --    &   --    &   --   \\      %Pipe-29a
 70\,\,  &  &   0.14  &   3.82  &   0.23  &   0.08  &  &   --    &   --    &   --    &   --   \\      %Pipe-24a
 74\,\,  &  &   --    &   --    &   --    &   --    &  &   --    &   --    &   --    &   --   \\      %Pipe-14a
 87*     &  &   2.28  &   4.47  &   0.14  &   0.75  &  &   0.29  &   4.46  &   0.16  &   0.11   \\      %Pipe-2a
 89\,\,  &  &   0.09  &   4.45  &   0.10  &   0.02  &  &   --    &   --    &   --    &   --   \\      %Pipe-11a
 91\,\,  &  &   0.17  &   4.21  &   0.07  &   0.03  &  &   --    &   --    &   --    &   --   \\      %Pipe-23a
 92\,\,  &  &   0.17  &   5.10  &   0.19  &   0.08  &  &   --    &   --    &   --    &   --   \\      %Pipe-8a
 93\,\,  &  &   0.43  &   5.17  &   0.17  &   0.18  &  &   --    &   --    &   --    &   --   \\      %Pipe-7a
 97\,\,  &  &   0.39  &   3.81  &   0.21  &   0.19  &  &   --    &   --    &   --    &   --   \\      %Pipe-9a
 99\,\,  &  &   --    &   --    &   --    &   --    &  &   --    &   --    &   --    &   --   \\      %Pipe-85a
101*     &  &   1.06  &   3.31  &   0.09  &   0.22  &  &   0.22  &   3.33  &   0.09  &   0.05   \\      %Pipe-13a
102\,\,  &  &   0.14  &   4.87  &   0.24  &   0.08  &  &   --    &   --    &   --    &   --   \\      %Pipe-4a
108*     &  &   0.31  &   3.22  &   0.16  &   0.12  &  &   --    &   --    &   --    &   --   \\      %Pipe-83
109*     &  &   3.90  &   5.75  &   0.08  &   0.73  &  &   1.22  &   5.73  &   0.08  &   0.22   \\      %Pipe-6
113\,\,  &  &   0.16  &   4.60  &   0.06  &   0.02  &  &   --    &   --    &   --    &   --   \\      %Pipe-49a
132\,\,  &  &   0.20  &   3.91  &   0.18  &   0.08  &  &   --    &   --    &   --    &   --   \\      %Pipe-26
\tableline
\end{tabular}
\tablenotetext{a}{Asterisks mark cores for which the \ammoniaoneone\, \vlsr\, and \sigmav\, were determined via the forward-fitting routine.}
\end{center}}}
\end{table}

%%%%%%%%%%%%%%%%%%%%%%%%%%%%%%%%%%%%%%%%%%%%%%%%%%%%%%%%%%%%%%%%%%%%%%%%%%%%%%%%%%%%

\begin{table}
{\scriptsize{
\begin{center}
\caption{\label{line-parameters-2}Observed line parameters of \ccs\, and \hcfiven.}
\begin{tabular}{rcccccccccc}
\tableline  \tableline
Core & &  \multicolumn{4}{c}{CCS} & &\multicolumn{4}{c}{\hcfivennt} \\  
\cline{2-6} \cline{8-11}
     & &\tastar\, & \vlsr & \sigmav & I & &\tastar\, & \vlsr & \sigmav & I \\
     & & (K) & (\kms) & (\kms) & (\Kkms) & & (K)  &(\kms) & (\kms) & (\Kkms)  \\
\tableline
  6  &  &  0.23  &   3.50  &   0.07  &   0.03  &  &   --  &   --  &   --  &   --   \\      %Pipe-12a
  7  &  &   --  &   --  &   --  &   --  &  &   --  &   --  &   --  &   --   \\      %Pipe-17a
  8  &  &   --  &   --  &   --  &   --  &  &   --  &   --  &   --  &   --   \\      %Pipe-15a
 11  &  &   --  &   --  &   --  &   --  &  &   --  &   --  &   --  &   --   \\      %Pipe-22a
 12  &  &   0.35  &   3.47  &   0.19  &   0.15  &  &   0.58  &   3.44  &   0.21  &   0.29   \\      %Pipe-B59c-2
 13  &  &   --  &   --  &   --  &   --  &  &   --  &   --  &   --  &   --   \\      %Pipe-131
 14  &  &   0.14  &   3.45  &   0.09  &   0.03  &  &   --  &   --  &   --  &   --   \\      %Pipe-5
 15  &  &   --  &   --  &   --  &   --  &  &   --  &   --  &   --  &   --   \\      %Pipe-33a
 17  &  &   --  &   --  &   --  &   --  &  &   --  &   --  &   --  &   --   \\      %Pipe-129
 20  &  &   --  &   --  &   --  &   --  &  &   --  &   --  &   --  &   --   \\      %Pipe-44a
 22  &  &   --  &   --  &   --  &   --  &  &   --  &   --  &   --  &   --   \\      %Pipe-73a
 23  &  &   --  &   --  &   --  &   --  &  &   --  &   --  &   --  &   --   \\      %Pipe-58a
 25  &  &   --  &   --  &   --  &   --  &  &   --  &   --  &   --  &   --   \\      %Pipe-104a
 27  &  &   --  &   --  &   --  &   --  &  &   --  &   --  &   --  &   --   \\      %Pipe-102a
 30  &  &   --  &   --  &   --  &   --  &  &   --  &   --  &   --  &   --   \\      %Pipe-128
 31  &  &   --  &   --  &   --  &   --  &  &   --  &   --  &   --  &   --   \\      %Pipe-101a
 33  &  &   --  &   --  &   --  &   --  &  &   --  &   --  &   --  &   --   \\      %Pipe-31a
 34  &  &   --  &   --  &   --  &   --  &  &   --  &   --  &   --  &   --   \\      %Pipe-61a
 36  &  &   --  &   --  &   --  &   --  &  &   --  &   --  &   --  &   --   \\      %Pipe-105
 37  &  &   0.14  &   3.32  &   0.08  &   0.03  &  &   --  &   --  &   --  &   --   \\      %Pipe-56a
 40  &  &   0.78  &   3.35  &   0.06  &   0.11  &  &   0.30  &   3.34  &   0.14  &   0.10   \\      %Pipe-1
 41  &  &   --  &   --  &   --  &   --  &  &   --  &   --  &   --  &   --   \\      %Pipe-40a
 42  &  &   0.07  &   3.86  &   0.11  &   0.02  &  &   --  &   --  &   --  &   --   \\      %Pipe-3
 47  &  &   --  &   --  &   --  &   --  &  &   --  &   --  &   --  &   --   \\      %Pipe-34a
 48  &  &   --  &   --  &   --  &   --  &  &   --  &   --  &   --  &   --   \\      %Pipe-27a
 51  &  &   --  &   --  &   --  &   --  &  &   --  &   --  &   --  &   --   \\      %Pipe-57a
 56  &  &   --  &   --  &   --  &   --  &  &   --  &   --  &   --  &   --   \\      %Pipe-21a
 61  &  &   --  &   --  &   --  &   --  &  &   --  &   --  &   --  &   --   \\      %Pipe-52a
 62  &  &   --  &   --  &   --  &   --  &  &   --  &   --  &   --  &   --   \\      %Pipe-42a
 65  &  &   --  &   --  &   --  &   --  &  &   --  &   --  &   --  &   --   \\      %Pipe-20a
 66  &  &   --  &   --  &   --  &   --  &  &   --  &   --  &   --  &   --   \\      %Pipe-29a
 70  &  &   --  &   --  &   --  &   --  &  &   --  &   --  &   --  &   --   \\      %Pipe-24a
 74  &  &   --  &   --  &   --  &   --  &  &   --  &   --  &   --  &   --   \\      %Pipe-14a
 87  &  &   0.23  &   4.50  &   0.25  &   0.14  &  &   --  &   --  &   --  &   --   \\      %Pipe-2a
 89  &  &   0.06  &   4.87  &   0.26  &   0.04  &  &   --  &   --  &   --  &   --   \\      %Pipe-11a
 91  &  &   --  &   --  &   --  &   --  &  &   --  &   --  &   --  &   --   \\      %Pipe-23a
 92  &  &   0.15  &   5.04  &   0.12  &   0.04  &  &   --  &   --  &   --  &   --   \\      %Pipe-8a
 93  &  &   0.16  &   5.26  &   0.17  &   0.06  &  &   --  &   --  &   --  &   --   \\      %Pipe-7a
 97  &  &   --  &   --  &   --  &   --  &  &   --  &   --  &   --  &   --   \\      %Pipe-9a
 99  &  &   --  &   --  &   --  &   --  &  &   --  &   --  &   --  &   --   \\      %Pipe-85a
101  &  &   0.55  &   3.37  &   0.07  &   0.09  &  &   0.44  &   3.36  &   0.15  &   0.15   \\      %Pipe-13a
102  &  &   0.35  &   4.84  &   0.08  &   0.07  &  &   --  &   --  &   --  &   --   \\      %Pipe-4a
108  &  &   --  &   --  &   --  &   --  &  &   --  &   --  &   --  &   --   \\      %Pipe-83
109  &  &   0.44  &   5.80  &   0.07  &   0.07  &  &   0.27  &   5.76  &   0.12  &   0.08   \\      %Pipe-6
113  &  &   --  &   --  &   --  &   --  &  &   --  &   --  &   --  &   --   \\      %Pipe-49a
132  &  &   --  &   --  &   --  &   --  &  &   --  &   --  &   --  &   --   \\      %Pipe-26
\tableline
\end{tabular}
\end{center}}}
\end{table}

%%%%%%%%%%%%%%%%%%%%%%%%%%%%%%%%%%%%%%%%%%%%%%%%%%%%%%%%%%%%%%%%%%%%%%%%%%%%%%%%%%%%

\begin{table}
\begin{center}
\caption{\label{fitspec}Derived parameters from the \ammonia\, lines.}
\begin{tabular}{crrrr}
\tableline  \tableline
Core & \multicolumn{1}{c}{\tk}  & \multicolumn{1}{c}{$\tau$}  & \multicolumn{1}{c}{\tex} &  \multicolumn{1}{c}{\tex $\tau$} \\%& $\Delta$V$_{th}$ & $\Delta$V & ratio & $\Delta_{non-th}$  \\
     & \multicolumn{1}{c}{(K)}  &                             & \multicolumn{1}{c}{(K)}  & \multicolumn{1}{c}{(K)} \\%& (\kms) & (\kms)&$\Delta$V/$\Delta$V$_{th}$  & (\kms)\\
\tableline

 12  &  11.3 $\pm$  0.1 &10.2 $\pm$  0.2         & 7.4 $\pm$  0.0         & \multicolumn{1}{c}{--} \\%&   0.17  &   0.35  &   2.0  &  0.31 \\%Pipe-B59c-2
 14  &  12.0 $\pm$  0.4 & 1.0 $\pm$  0.3         & 4.4 $\pm$  0.4         & \multicolumn{1}{c}{--} \\%&   0.18  &   0.33  &   1.8  &  0.28   \\ %Pipe-5
 17  &  17.5 $\pm$  1.3 & \multicolumn{1}{c}{--} & \multicolumn{1}{c}{--} & 0.5 $\pm$  0.0         \\%&   0.22  &   0.59  &   2.7  &  0.55   \\%Pipe-129
 20  &  15.2 $\pm$  0.5 & \multicolumn{1}{c}{--} & \multicolumn{1}{c}{--} & 1.7 $\pm$  0.1         \\%&   0.20  &   0.40  &   2.0  &  0.34   \\%Pipe-44a
 40  &  10.3 $\pm$  0.5 & 2.4 $\pm$  0.4         & 5.3 $\pm$  0.3         & \multicolumn{1}{c}{--} \\%&   0.17  &   0.23  &   1.4  &  0.17   \\ %Pipe-1
 41  &  14.3 $\pm$  0.7 & \multicolumn{1}{c}{--} & \multicolumn{1}{c}{--} & 1.6 $\pm$  0.1         \\%&   0.20  &   0.28  &   1.4  &  0.20   \\%Pipe-40a
 42  &  12.4 $\pm$  0.4 & \multicolumn{1}{c}{--} & \multicolumn{1}{c}{--} & 1.9 $\pm$  0.0         \\%&   0.18  &   0.26  &   1.4  &  0.18   \\ %Pipe-3
 47  &  12.6 $\pm$  1.0 & 2.5 $\pm$  0.8         & 3.2 $\pm$  0.1         & \multicolumn{1}{c}{--} \\%&   0.18  &   0.33  &   1.8  &  0.27   \\%Pipe-34a
 87  &   9.8 $\pm$  0.3 & 6.3 $\pm$  0.3         & 5.4 $\pm$  0.1         & \multicolumn{1}{c}{--} \\%&   0.16  &   0.33  &   2.0  &  0.29   \\ %Pipe-2a
101  &  10.4 $\pm$  0.2 & 5.1 $\pm$  0.2         & 4.5 $\pm$  0.1         & \multicolumn{1}{c}{--} \\%&   0.17  &   0.21  &   1.3  &  0.13   \\%Pipe-13a
108  &  14.1 $\pm$  0.8 & \multicolumn{1}{c}{--} & \multicolumn{1}{c}{--} & 1.0 $\pm$  0.0         \\%&   0.19  &   0.38  &   1.9  &  0.32   \\ %Pipe-83
109  &   9.5 $\pm$  0.1 &19.8 $\pm$  0.3         & 6.6 $\pm$  0.0         & \multicolumn{1}{c}{--} \\%&   0.16  &   0.19  &   1.2  &  0.10   \\ %Pipe-6
\tableline
\end{tabular}
\end{center}
\end{table}

%%%%%%%%%%%%%%%%%%%%%%%%%%%%%%%%%%%%%%%%%%%%%%%%%%%%%%%%%%%%%%%%%%%%%%%%%%%%%%%%%%%%

\begin{table}
\begin{center}
\caption{\label{column-densities}Volume and column densities of the brightest cores in the Pipe Nebula.}
\begin{tabular}{cccccc}
\tableline  \tableline
Core & log[n(\hh)]\tablenotemark{a} & log[N(\ammonia)] & log[N(CCS)] & log[N(\hcfivennt)] & log[N(\ceont)] \\
     & (cm$^{-3}$) & (cm$^{-2}$)      & (cm$^{-2}$) & (cm$^{-2}$) & (cm$^{-2}$) \\
\tableline
  6    &   --  &   13.7  &   12.4  &   --  &   15.1 \\    %Pipe-12a
  7    &   --  &   13.2  &   --  &   --  &   15.1 \\    %Pipe-17a
  8    &   --  &   13.3  &   --  &   --  &   15.1 \\    %Pipe-15a
 12    &    4.4  &   15.2  &   13.1  &   12.9  &   -- \\    %Pipe-B59c-2
 14    &    3.7  &   14.0  &   12.4  &   --  &   15.3 \\    %Pipe-5
 15    &   --  &   13.4  &   --  &   --  &   15.2 \\    %Pipe-33a
 17    &   --  &   13.3  &   --  &   --  &   14.5 \\    %Pipe-129
 20    &   --  &   13.6  &   --  &   --  &   14.9 \\    %Pipe-44a
 22    &   --  &   13.2  &   --  &   --  &   14.7 \\    %Pipe-73a
 23    &   --  &   13.0  &   --  &   --  &   14.8 \\    %Pipe-58a
 25    &   --  &   13.3  &   --  &   --  &   14.5 \\    %Pipe-104a
 37    &   --  &   --  &   12.3  &   --  &   15.0 \\    %Pipe-56a
 40    &    4.0  &   14.3  &   13.0  &   12.4  &   15.1 \\    %Pipe-1
 41    &   --  &   13.4  &   --  &   --  &   15.0 \\    %Pipe-40a
 42    &   --  &   13.5  &   12.2  &   --  &   15.2 \\    %Pipe-3
 47    &    3.0  &   14.2  &   --  &   --  &   15.1 \\    %Pipe-34a
 65    &   --  &   13.6  &   --  &   --  &   15.2 \\    %Pipe-20a
 70    &   --  &   13.3  &   --  &   --  &   15.2 \\    %Pipe-24a
 87    &    4.0  &   14.8  &   13.0  &   --  &   15.1 \\    %Pipe-2a
 89    &   --  &   13.1  &   12.5  &   --  &   15.1 \\    %Pipe-11a
 91    &   --  &   13.2  &   --  &   --  &   14.9 \\    %Pipe-23a
 92    &   --  &   13.4  &   12.5  &   --  &   15.2 \\    %Pipe-8a
 93    &   --  &   13.8  &   12.7  &   --  &   15.2 \\    %Pipe-7a
 97    &   --  &   13.8  &   --  &   --  &   14.8 \\    %Pipe-9a
101    &    3.7  &   14.5  &   12.9  &   12.6  &   14.9 \\    %Pipe-13a
102    &   --  &   13.4  &   12.7  &   --  &   14.9 \\    %Pipe-4a
108    &   --  &   13.4  &   --  &   --  &   14.6 \\    %Pipe-83
109    &    4.4  &   15.2  &   12.7  &   12.3  &   15.2 \\    %Pipe-6
113    &   --  &   13.2  &   --  &   --  &   14.9 \\    %Pipe-49a
132    &   --  &   13.4  &   --  &   --  &   15.1 \\    %Pipe-26
\tableline
\end{tabular}
\tablenotetext{a}{Calculated from the \ammonia\, data as described in \S~\ref{ammonia-nhh}.}
\end{center}
\end{table}

%%%%%%%%%%%%%%%%%%%%%%%%%%%%%%%%%%%%%%%%%%%%%%%%%%%%%%%%%%%%%%%%%%%%%%%%%%%%%%%%%%%%

%%%%%%%%%%%%%%%%%%%%%%%%%%%%%%%%%%%%%%%%%%%%%%%%%%%%%%%%%%%%%%%%%%%%%%%%%%%%%%%%%%%%
% Figures
%%%%%%%%%%%%%%%%%%%%%%%%%%%%%%%%%%%%%%%%%%%%%%%%%%%%%%%%%%%%%%%%%%%%%%%%%%%%%%%%%%%%

%%%%%%%%%%%%%%%%%%%%%%%%%%%%%%%%%%%%%%%%%%%%%%%%%%%%%%%%%%%%%%%%%%%%%%%%%%%%%%%%%%%%
\clearpage
%%%%%%%%%%%%%%%%%%%%%%%%%%%%%%%%%%%%%%%%%%%%%%%%%%%%%%%%%%%%%%%%%%%%%%%%%%%%%%%%%%%%

%%%%%%%%%%%%%%%%%%%%%%%%%%%%%%%%%%%%%%%%%%%%%%%%%%%%%%%%%%%%%%%%%%%%%%%%%%%%%%%%%%%%
\clearpage
%%%%%%%%%%%%%%%%%%%%%%%%%%%%%%%%%%%%%%%%%%%%%%%%%%%%%%%%%%%%%%%%%%%%%%%%%%%%%%%%%%%%

%%%%%%%%%%%%%%%%%%%%%%%%%%%%%%%%%%%%%%%%%%%%%%%%%%%%%%%%%%%%%%%%%%%%%%%%%%%%%%%%%%%%
\clearpage
%%%%%%%%%%%%%%%%%%%%%%%%%%%%%%%%%%%%%%%%%%%%%%%%%%%%%%%%%%%%%%%%%%%%%%%%%%%%%%%%%%%%

%%%%%%%%%%%%%%%%%%%%%%%%%%%%%%%%%%%%%%%%%%%%%%%%%%%%%%%%%%%%%%%%%%%%%%%%%%%%%%%%%%%%
\clearpage
%%%%%%%%%%%%%%%%%%%%%%%%%%%%%%%%%%%%%%%%%%%%%%%%%%%%%%%%%%%%%%%%%%%%%%%%%%%%%%%%%%%%

%%%%%%%%%%%%%%%%%%%%%%%%%%%%%%%%%%%%%%%%%%%%%%%%%%%%%%%%%%%%%%%%%%%%%%%%%%%%%%%%%%%%
\clearpage
%%%%%%%%%%%%%%%%%%%%%%%%%%%%%%%%%%%%%%%%%%%%%%%%%%%%%%%%%%%%%%%%%%%%%%%%%%%%%%%%%%%%

%%%%%%%%%%%%%%%%%%%%%%%%%%%%%%%%%%%%%%%%%%%%%%%%%%%%%%%%%%%%%%%%%%%%%%%%%%%%%%%%%%%%
\clearpage
%%%%%%%%%%%%%%%%%%%%%%%%%%%%%%%%%%%%%%%%%%%%%%%%%%%%%%%%%%%%%%%%%%%%%%%%%%%%%%%%%%%%

%%%%%%%%%%%%%%%%%%%%%%%%%%%%%%%%%%%%%%%%%%%%%%%%%%%%%%%%%%%%%%%%%%%%%%%%%%%%%%%%%%%%
\clearpage
%%%%%%%%%%%%%%%%%%%%%%%%%%%%%%%%%%%%%%%%%%%%%%%%%%%%%%%%%%%%%%%%%%%%%%%%%%%%%%%%%%%%

%%%%%%%%%%%%%%%%%%%%%%%%%%%%%%%%%%%%%%%%%%%%%%%%%%%%%%%%%%%%%%%%%%%%%%%%%%%%%%%%%%%%
\clearpage
%%%%%%%%%%%%%%%%%%%%%%%%%%%%%%%%%%%%%%%%%%%%%%%%%%%%%%%%%%%%%%%%%%%%%%%%%%%%%%%%%%%%

%%%%%%%%%%%%%%%%%%%%%%%%%%%%%%%%%%%%%%%%%%%%%%%%%%%%%%%%%%%%%%%%%%%%%%%%%%%%%%%%%%%%
\clearpage
%%%%%%%%%%%%%%%%%%%%%%%%%%%%%%%%%%%%%%%%%%%%%%%%%%%%%%%%%%%%%%%%%%%%%%%%%%%%%%%%%%%%

%%%%%%%%%%%%%%%%%%%%%%%%%%%%%%%%%%%%%%%%%%%%%%%%%%%%%%%%%%%%%%%%%%%%%%%%%%%%%%%%%%%%
% APPENDIX
%%%%%%%%%%%%%%%%%%%%%%%%%%%%%%%%%%%%%%%%%%%%%%%%%%%%%%%%%%%%%%%%%%%%%%%%%%%%%%%%%%%%

\clearpage

\end{document}